\documentclass{aastex}
\usepackage{natbib,graphicx,latexsym}
\usepackage{emulateapj5}

\newcommand{\HST}{{\sl HST}}

\newcommand{\perone}{\mbox{$^{-1}$}}
\newcommand{\pertwo}{\mbox{$^{-2}$}}

\newcommand{\etal}{et al.}
\newcommand{\eg}{e.g.}
\newcommand{\ie}{i.e.}

\newcommand{\kms}{\hbox{km~s$^{-1}$}}
\newcommand{\rchisq}{\mbox{$\tilde\chi^2$}}
\newcommand{\Ho}{\mbox{$H_0$}}
\newcommand{\Zsun}{\mbox{$Z_{\odot}$}}

\newcommand{\Ic}{\mbox{${I_c}$}}

\newcommand{\Ks}{\mbox{$K_S$}}
\newcommand{\Kp}{\mbox{$K^{\prime}$}}

\newcommand{\VmI}{\mbox{$V\!-\!I_c$}}

\newcommand{\Vbar}{\mbox{$\overline{V}$}}
\newcommand{\Rbar}{\mbox{$\overline{R_c}$}}
\newcommand{\Ibar}{\mbox{$\overline{I_c}$}}

\newcommand{\Ksbar}{\mbox{$\overline{K_S}$}}

\newcommand{\Mbar}{\mbox{$\overline{M}$}}
\newcommand{\mbar}{\mbox{$\overline{m}$}}

\newcommand{\MbarKs}{\mbox{$\overline{M}_{K_S}$}}
\newcommand{\mbarKs}{\mbox{$\overline{m}_{K_S}$}}

\slugcomment{{\em Astrophysical Journal}, in press}

\shorttitle{Liu, Graham, \& Charlot}
\shortauthors{IR SBFs of Fornax Cluster Galaxies}

\begin{document}

\title{Surface Brightness Fluctuations of Fornax Cluster Galaxies:\\
Calibration of Infrared SBFs and Evidence for Recent Star Formation}

\author{\sc Michael C. Liu\altaffilmark{1,2} and James R. Graham}
\affil{Astronomy Department, University of California, Berkeley, CA
94720} 

\altaffiltext{1}{Visiting Astronomer, Cerro Tololo Interamerican
Observatory (CTIO), National Optical Astronomy Observatories, which is
operated by Associated Universities for Research in Astronomy, Inc.,
under cooperative agreement with the National Science Foundation.}

\altaffiltext{2}{Currently Beatrice Watson Parrent Fellow at the
Institute for Astronomy, University of Hawai`i, 2680 Woodlawn Drive,
Honolulu, HI 96822.}  

\email{mliu@ifa.hawaii.edu}

\author{\sc St\'ephane Charlot\altaffilmark{3}}
\affil{Institut d'Astrophysique de Paris, CNRS, 75014 Paris, France} 
\altaffiltext{3}{Also Max-Planck Institut f\"ur Astrophysik,
Karl-Schwarzschild-Strasse 1, 85748, Garching, Germany}

\begin{abstract}

We have measured $K_S$-band (2.0--2.3~\micron) surface brightness
fluctuations (SBFs) of 19~early-type galaxies in the Fornax cluster.
Fornax is ideally suited both for calibrating SBFs as distance
indicators and for using SBFs to probe the unresolved stellar content of
early-type galaxies.  Combining our results with published data for
other nearby clusters, we calibrate $K_S$-band SBFs using \HST\ Cepheid
cluster distances and $I$-band SBF distances to individual galaxies.
With the latter, the resulting calibration is
$$ \MbarKs = (-5.84\pm0.04) + (3.6\pm0.8) [(\VmI)_0 - 1.15], $$
valid for \hbox{$1.05 < (\VmI)_0 < 1.25$} and not including any
systematic errors in the \HST\ Cepheid distance scale.  The fit accounts
for the covariance between \VmI\ and \MbarKs\ when calibrated in this
fashion. The intrinsic cosmic scatter of \MbarKs\ appears to be larger
than that of $I$-band SBFs.  S0 galaxies may follow a different relation
though the data are inconclusive.  The discovery of a correlation between
\Ks-band fluctuation magnitudes and colors with \VmI\ is a new clue into
the star formation histories of early-type galaxies.  This relation
naturally accounts for galaxies previously claimed to have anomalously
bright $K$-band SBFs, namely M32 and NGC~4489. Models indicate that the
stellar populations dominating the SBF signal have a significant range
in age; some scatter in metallicity may also be present.  The youngest
ages imply some galaxies have very luminous giant branches, akin to
those in intermediate-age (few Gyr) Magellanic Cloud clusters.  The
inferred metallicities are roughly solar, though this depends on the
choice of theoretical models.  A few Fornax galaxies have unusually
bright \Ks-band SBFs, perhaps originating from a high metallicity burst
of star formation in the last few~Gyr.  The increased spread and
brightening of the \Ks-band SBFs with bluer \VmI\ suggest that the lower
mass cluster galaxies ($\lesssim 0.1 L^*$) may have had more extended
and more heterogeneous star formation histories than those of the more
massive galaxies.

\end{abstract}



\section{Introduction}

Surface brightness fluctuations (SBFs) provide a powerful method to
determine distances to early-type galaxies \citep{1988AJ.....96..807T}.
Unlike many cosmological distance indicators, this method has a firm
physical basis: SBFs arise from Poisson fluctuations in the number of
stars within a resolution element.  This method uses the fact that the
ratio of the second moment of the stellar luminosity function (LF) to
the first moment of the stellar LF has units of luminosity.  (See
reviews by \citealt{1992PASP..104..599J}, \citealt{1997eds..proc..297T},
and \citealt{1999phcc.conf..181B}.)  Using SBFs as a distance indicator
requires that the bright end of the stellar LF in elliptical galaxies
and spiral bulges is universal, or that variations in the LF from galaxy
to galaxy can be measured and calibrated.  Tonry and collaborators have
extensively pursued $I$-band SBF measurements
\citep[e.g.][]{2000ApJ...530..625T,sbf4}.  They have found $I$-band SBFs
vary strongly between galaxies, but these variations are well correlated
with \VmI\ galaxy color.

Much less work has been done on SBF measurements in the near-infrared
(1--2.5~\micron).  Since SBFs in ellipticals are dominated by late-type
giant stars, SBFs are brightest in the IR, with SBFs in $K$-band
(2.2~\micron) about 4~mag brighter than in $I$-band
\citep{1993ApJ...410...81L}.  The color of the night sky is even redder
($I-K\gtrsim5$) meaning IR observations need longer integrations to
achieve the same S/N.  However, the extreme redness of SBFs ($V$--$K
\approx 7$) compared to contaminating globular clusters ($V$--$K \approx
2.5$) means that SBFs can ultimately be measured in the IR to greater
distances than in the optical, out to $cz \approx 10,000$~km/s using
8--10 meter class telescopes \citep{liu2001} or {\sl Hubble Space
Telescope} (\HST; \citealp{jen01}), and thus determine $H_0$ at
distances where peculiar velocities are negligible.  IR measurements are
also less subject to uncertainties in dust and extinction corrections
than optical ones. Moreover, if IR SBFs can be shown to have a small
intrinsic dispersion, they would be a potent tool for peculiar velocity
studies out to large distances.

As with any distance indicator, the utility of IR SBFs relies on the
accuracy of their zero point and a thorough understanding of their
precision.  Published IR measurements to date
\citep{1993ApJ...410...81L, 1994ApJ...433..567P, 1996ApJ...468..519J,
1998ApJ...505..111J, 1999ApJ...510...71J, mei2001} have all been in the
$K$-band.  The current calibration of $K$-band SBFs rests on the work of
\citet{1998ApJ...505..111J}. They establish $K$-SBF as a secondary
distance indicator using observations of M31 and four~Virgo cluster
galaxies, with a Virgo distance from \HST\ Cepheid observations. They
also use $I$-band SBF distances to derive a tertiary calibration for
$K$-band SBFs from a larger set of 11 galaxies.

Since SBFs depend on the second moment of the stellar luminosity
function, they also provide data on unresolved stellar populations
unattainable from the first moment alone, i.e. integrated galaxian light
and spectra.  For instance, IR SBFs are sensitive to the presence of
intermediate age ($t\lesssim5$~Gyr) asymptotic giant branch (AGB)
stars. The SBF signal in the optical and IR is dominated by stars on the
red (first ascent) giant branch and asymptotic giant branch.  $I$-band
SBFs are predicted by models to be roughly degenerate to age and
metallicity changes \citep{1993ApJ...409..530W, bc2000sbf}, hence their
utility as a distance indicator and inadequacy for stellar population
studies. On the other hand, IR SBF magnitudes (and optical/IR SBF
colors) are predicted to show strong variations with age and
metallicity. This can potentially offer insights into the star formation
history of early-type galaxies, unique from ordinary integrated light
and spectra.  (We refer to \citealp{bc2000sbf} for a full discussion of
stellar population studies using SBF measurements.)

In this paper we present results from a study aimed at (1)~accurately
calibrating IR SBFs for cosmological distance measurements and (2)~using
multicolor SBF measurements in concert with state-of-the-art stellar
population synthesis models to study the unresolved stellar populations
of early-type galaxies.  We have observed the \Ks-band SBFs of 19
early-type galaxies in the nearby Fornax cluster --- Fornax is an ideal
place to calibrate IR SBFs as it is the nearest cluster ($d\sim19$~Mpc)
that both is compact and has many ellipticals.  The \HST\ Key Project
has measured Cepheid distances to three galaxies in the cluster
\citep{1999ApJ...515....1S, 1999ApJ...515...29M, 1999ApJ...525...80P,
2000ApJ...528..655M}. Fornax's small angular extent implies that errors
due to cluster depth should be small, unlike the case for the comparably
nearby Virgo cluster \citep[e.g.,][]{1997ApJS..108..417Y,
2000ApJ...536..255N}.  In addition, by observing galaxies with a spread
in properties (e.g., mean color), we can also measure any second
parameter effects and the intrinsic IR SBF dispersion.

In \S~\ref{sec:obs} we describe our observations and basic
reductions. Our methods for measuring the stellar SBF signal are
presented in \S~\ref{sec:measurements}, and our results are presented in
\S~\ref{sec:results}. In \S~\ref{sec:analysis}, we present the resulting
calibration of the \Ks-band SBFs and discuss the stellar populations of
the cluster galaxies, including evidence for recent star formation.  For
readers primarily interested in the results, we suggest skipping to
\S~\ref{sec:results} and \ref{sec:analysis}.  The results presented here
supersede the preliminary versions published in \citet{liu99b} and
\citet{liu99c}.


\section{Observations} \label{sec:obs}

\subsection{Blanco 4-m: SBF Data}

We observed Fornax cluster galaxies using the Blanco 4-m telescope at
Cerro Tololo Inter-American Observatory (CTIO) on 1997
November~08--11~UT and 1998 November~11--13~UT. We used the facility
near-IR camera CIRIM with the \Ks-band filter
\citep[2.0--2.3~\micron;][]{1995ApJS...96..117M}.  The camera employs a
Rockwell NICMOS-3 $256\!\times\!256$ HgCdTe array with a pixel scale of
0\farcs419~pixel\perone\ (\S~\ref{sec:surfphot}) when used with the f/8
secondary.  Both runs were photometric, except for parts of the very
first night and the end of the very last night. The typical seeing FWHM
was 0\farcs8, with a full range of 0\farcs6 to 1\farcs1.

Our sample is listed in Table~\ref{table-gals}.  Integrations on each
galaxy were interlaced with blank sky fields in pairs, with data taken
in an ABBA pattern, starting with the sky field.  Offsets from galaxy to
sky fields were typically several arcminutes, with larger offsets for
brighter galaxies.  The sky fields were chosen to be free of bright
stars and away from other large galaxies.
For both the galaxy and sky positions, the center of each field was
ultimately dithered in a $2\times2$ square pattern over the detector
which was then repeated.  An off-axis CCD camera mounted on a motorized
X/Y stage was used to guide the telescope during the integrations.  When
moving between galaxy and sky fields, the stage was moved to maintain
the position of the guide star on the CCD camera so that the same star
was used for both fields.  Mindful of the evidence that low S/N can lead
to large systematic errors in SBF measurements
\citep{1996ApJ...468..519J}, we made sure to acquire plenty of
integration.  The total integration on each galaxy was typically 32~min,
with an equal amount of integration on the blank sky field.

\subsection{CTIO 1.5-m: Surface Photometry}

Because most of our targets fill the field of view of CIRIM on the
Blanco 4-m telescope, we also obtained wider-field images using CIRIM on
the CTIO 1.5-m telescope. CIRIM on the 1.5-m telescope with the f/8
secondary has a plate scale of 1\farcs14~pixel\perone\ with the \Ks\
filter.  Data were obtained in a similar fashion as for the 4-m data,
using interlaced pairs of sky and galaxy fields.  The 1.5-m observations
were not guided. The galaxy was stepped in a $2\times2$ pattern over the
detector to cover a final area of
\hbox{$\approx8\arcmin\!\times\!8\arcmin$} centered on each galaxy.
Offsets between galaxy and sky fields were $\approx
8\arcmin-10\arcmin$. The total integration on the center of each galaxy
was typically 16~min.

Our first 1.5-m run was 1997 October~11--12~UT. Conditions were
photometric for the second half of the first night and most of the
second night.  Galaxies observed during non-photometric conditions on
the first night were re-observed on the second night during photometric
conditions.  The atmospheric seeing was variable, ranging from
1\farcs5--2\farcs0~FWHM.  Our second 1.5-m run was 1998
November~08--10~UT, and conditions were photometric for all the data
presented herein. The seeing was variable, and the images have an
angular resolution ranging of 1\farcs1--1\farcs5~FWHM.

\subsection{Reductions} \label{sec:redux}

The data were reduced in a mostly standard manner for near-IR images.
We took great care to ensure that the reduction process did not
introduce any spatial correlations on the scale of the point spread
function (PSF), which would contaminate the SBF signal and lead to
systematic errors. All reductions and analysis were done using custom
software written for Research System Incorporated's IDL software
package.


We constructed an average bias frame from the median average of many
dark frames for every combination of integration time and coadds.  This
average bias was then subtracted from all images with the same
combination of integration time and coadds.  We created a flat field for
each night from a series of dithered images of the twilight sky.  We
used the twilight sky as it is expected to be the best approximation in
the \Ks-band to a surface of uniform illumination. However, since this
bandpass also included thermal emission from objects at room
temperature, the total illumination on the detector may not have been
non-uniform due to, \eg, warm dust particles on the instrument entrance
window or scattered thermal emission from the telescope structure.
Therefore, a flat field created by direct averaging of the twilight sky
images would have systematic errors due to the non-uniform component of
the illumination.  Instead, we used an iterative-fitting scheme to solve
for the relative quantum efficiency (QE) of each pixel using the entire
set of images simultaneously.  For each pixel $(x,y)$ in frame $i$ of a
set of twilight images, we represented the measured counts $DN_i$ as
\begin{equation}
DN_i(x,y) = q(x,y) \times S_i + k(x,y)
\label{eqn:flatfield}
\end{equation}
where $q(x,y)$ is the relative QE of the pixel (\ie, the flat field),
and $S_i$ and $k(x,y)$ are the number of counts from the uniform and
non-uniform illumination components, respectively.  By definition, the
median values of $q$ and $k$ were 1 and 0, respectively, and $k(x,y)$
could be either positive or negative as it represented the deviations
from uniform illumination. Also by definition, $S_i$ depended only on
the frame number $i$, and since it came from the twilight sky, it was a
monotonic function of $i$ (decreasing for sunset twiflats and increasing
for sunrise twiflats). Note that we assumed $k(x,y)$ was
time-independent, a reasonable assumption given that the twilight sky
frames were taken over a period of only a few minutes.
Equation~\ref{eqn:flatfield} is linear, so using the median count level
as $S_i$, we found the best-fitting $q(x,y)$ and $k(x,y)$.  For each
pixel, we used a $\sigma$-clipping scheme to flag and mask any outlier
frames due to the presence of a cosmic ray or a star.  The result was
the flat field $q(x,y)$ and a map of the non-uniform illumination
$k(x,y)$, which was of order a few percent of the mean twilight sky
flux.

After flat-fielding, the individual images still did not have a flat
(uniform) background.  This was corrected by sky subtraction, which
served two purposes: (1) to remove (most of) the night sky flux, which
can be $\gtrsim$100$\times$ the mean galaxy surface brightness, and (2)
to remove non-uniform structure arising from flat-fielding errors and
the aforementioned thermal emission. Note that even if the flat fields
were perfect, sky subtraction would still have been a necessary step
because of (2).

A preliminary sky subtraction of the images was performed to identify
astronomical objects.  Then for each image, we constructed a running sky
frame from the average of prior and subsequent images of the blank
fields, excluding any of the identified astronomical objects from the
averaging.  All the individual images, both of the galaxy and of the
adjacent blank sky fields, were processed in this fashion.  

The sky emission changes temporally so the running sky frame needed to
be accurately scaled before being subtracted. The usual way to scale the
image is to multiply the local sky frame so that its median matches that
of the object frame. However, we could not use this scaling approach for
the galaxy images --- the scaling factor would be slightly too large
since it would be based on the sum of the night sky brightness and the
median galaxy surface brightness.  The result would be a slight
over-subtraction of the local sky frame. Instead, we scale the local sky
frames to the median counts in unsubtracted galaxy images excluding a
large circular region centered on the galaxy, typically 80--120 pixels
in radius.  For the largest galaxies, there was still be a very slight
over-subtraction, but this error was corrected when we calibrated the DC
sky level (\S~\ref{sec:surfphot}).  We used these same circular masks
for the scaled sky subtraction of the blank sky images. This ensured the
blank sky images and galaxy images were reduced in basically an
identical fashion, which was useful for the SBF power spectrum analysis
(\S~\ref{sec:pspecfit}).

We used the center of the galaxy to register the individual frames,
which were then shifted by integer pixel offsets and averaged to
assemble a mosaic of the field.  We did not employ any sub-pixel
interpolation to align the images, since any such scheme would have
introduced high spatial frequency correlations in the pixel-to-pixel
noise and contaminated the SBF measurements. Bad pixels were identified
by inter-comparing the registered individual images to find outlier
pixels. These were then masked during the construction of the final
mosaic.

We observed the faint HST IR standards of \citet{1998AJ....116.2475P} as
flux calibrators. Hence, our resulting magnitudes are Vega-based.  On
most nights, we observed standards over a range of airmasses to measure
the zero point and extinction correction for each individual night.  On
some nights, it was not possible to observe standards over a wide enough
range of airmass; in this case, we used the zero point from standards
observed close in time and airmass to the galaxies and assumed an
extinction of 0.090~mag~airmass\perone, typical for our other nights.
Extinction corrections were obtained from the the DIRBE/IRAS-derived
values of \citet{1998ApJ...500..525S}.

\subsection{Calibration of Surface Photometry}
\label{sec:surfphot}

The final step in the image processing was to photometrically calibrate
the CTIO~4-m \Ks-band galaxy images used for SBF measurements. We used
the wider-field 1.5-m images to account for the DC sky background.
First we extracted azimuthally averaged surface brightness profiles from
the 4-m and 1.5-m images.  In extracting the profile from the 1.5-m
images, we used a 20\arcsec\ wide annulus located at the edge of the
images, typically 3\arcmin\ in radius, to assess the sky level.  We then
compared the 1.5-m and 4-m profiles and simultaneously fit for both a
multiplicative and additive offset between the two.  The multiplicative
term represented any zero point offset and the additive term was the sky
background.  In order to do this profile fitting accurately, we
carefully calibrated the plate scale of the 1.5-m and 4-m data relative
to each other, which was necessary for this profile matching
scheme. (The absolute plate scale for the 1.5-m data was determined
using stars in common with the Digitized Sky Survey.)  The differences
between the 4-m and 1.5-m photometric calibrations were quite small,
typically $\lesssim1\%$, providing confidence in the independent
calibrations.  The fitted sky level was then subtracted from the 4-m
images.

We checked the accuracy of our photometric calibrations against the
aperture photometry of \citet{1979ApJS...39...61P}. For the six galaxies
common to our two samples (NGC~1344, 1379, 1380, 1387, 1399, and 1404),
our photometry is on average 0.048~mag brighter than the Persson \etal\
large aperture (56\arcsec\ diameter) measurements, with a standard
deviation of 0.025~mag.  The Persson \etal\ data were taken with InSb
single-channel photometer and the CIT $K$-band filter, which has a
slightly redder bandpass (2.0--2.4~\micron) than our \Ks\ filter.  To
estimate the expected offset in the photometry, we synthesized
\Ks--$K_{CIT}$ colors using IR spectra of solar-metallicity M~giants
(spectral type M0--M5~III) compiled by \citet{1998PASP..110..863P},
which should be representative of elliptical galaxy spectra for this
purpose \citep{1978ApJ...220...75F}.  The synthesized \Ks--$K_{CIT}$
color is $-0.032\pm0.003$~mag, which includes accounting for the
filters' $k$-corrections appropriate for the redshift of the Fornax
galaxies. Therefore, we conclude the photometric calibrations of our
images are in good agreement with the Persson \etal\
measurements.\footnote{There is a larger discrepancy between our
photometry and the smaller aperture (29\arcsec\ diameter) photometry of
\citet{1979ApJS...39...61P}.  After accounting for the difference in
filter bandpasses along with the potential error from the centering of
the photometer aperture and the relatively smaller offsets to blank sky
positions, the remaining offset is $\approx\!0.1$~mag. Using curves of
growth measured from our images, the size of this offset suggests the
29\arcsec\ aperture of Persson \etal\ was in fact about 10\% larger in
diameter. Using a larger sample of galaxies, \citet{1999ApJS..124..127P}
has also suggested the smaller apertures of Persson \etal\ were actually
slightly larger than quoted.}


\section{SBF Measurements} \label{sec:measurements}

We measured SBF apparent magnitudes (\mbar) in a similar fashion to the
method described by \citet{1988AJ.....96..807T} and
\citet{1990AJ....100.1416T}.  The basic steps are: (1) measuring the
galaxy's mean surface brightness profile, (2) cataloguing the globular
clusters and background galaxies to quantify their contribution to the
fluctuations, (3) measuring the total fluctuation variance from the
Fourier power spectrum of the model-subtracted image, and (4)
determining \mbar\ by subtracting the variance due to contaminating
sources from the total measured variance.\footnote{When referring to
SBFs in this paper, we mean the fluctuations in the galaxy surface
brightness which arise from the statistical variations in the stellar
surface density, \ie, the stellar SBFs.  Sources unrelated to the galaxy
stellar population, unresolved globular clusters and faint background
galaxies, also produce fluctuations in the galaxy surface brightness
(see \S~\ref{varsub}).  On occasion, we refer to the fluctuations and/or
the variance due to these contaminating sources.  But the use of
``SBFs'' implicitly means the stellar SBFs.}  We now describe each of
these steps in detail.

\subsection{Surface Brightness Fitting}

Elliptical galaxy fits were done using the isophotal fitting scheme from
\citet{1979ApJ...227...56W}.  Contaminating point sources or galaxies
were first flagged and excluded from fitting.  Each isophote was
represented with an ellipse
\begin{equation}
x^2 + Sy^2 = A
\end{equation}
where $(x, y)$ are the coordinates for the ellipse in a reference frame
where the $x$-axis coincides with the major axis and the isophote's
center is at the origin. The quantity $S$ is the square of the ratio of
the major to the minor axis ($1/S = 1-e^2$, where $e$ is the
ellipticity), and $A$ is the square of the semi-major axis.  If $(x_0,
y_0)$ are the coordinates on the data frame, the transformation between
the two frames is
\begin{equation}
x = (y_0-y_c) \cos\alpha + (x_0-x_c) \sin\alpha
\end{equation}
\begin{equation}
y = (y_0-y_c) \sin\alpha - (x_0-x_c) \cos\alpha
\end{equation}
where $(x_c, y_c)$ is the location of the isophote center in the data
coordinates and $\alpha$ is the position angle of the major axis
measured clockwise from the $y_0$-axis.  For a given isophote with a
semi-major axis squared of $A_i$, we assumed the neighboring isophotes
had basically the same shape.  Thus, the only difference for different
pixels close to the isophote was the flux $F(x,y)$.  We assumed the flux
in the neighborhood of the isophote could be represented simply as
\begin{equation}
F(x,y) = F_i\ (A_i/A)^\gamma
\end{equation}
where $\gamma$ is the local slope of the intensity distribution.  

At each semi-major axis, we solved for the flux $F_i$, axial ratio
$\sqrt{S}$, position angle $\alpha$, center of the ellipse ($x_c, y_c$),
and radial dependence of the intensity profile $\gamma$.  The fits were
stepped outward logarithmically in radius, and all the parameters were
allowed to vary as a function of semi-major axis.  The fitted parameters
were then spline-interpolated as a function of semi-major axis, and a
galaxy model was constructed using bi-linear interpolation of the
isophotal parameters.

About half of the galaxies had obvious residuals from purely elliptical
isophotes.  In these cases, we also fit for the higher order harmonic
content of the isophotes.  The procedure we adopted is similar to the
widely used scheme of \citet{1987MNRAS.226..747J}.  At each semi-major
axis, we first solved for the ellipse parameters, as described above.
Then along this ellipse with flux $F_i$, we fitted the residuals as
\begin{equation}
F(E)-F_i = \sum_n A_n \sin(n E) + \sum_n\ B_n \cos(n E)
\end{equation}
where $E$ is the eccentric anomaly and $n\ge3$.  Typically the $n=3$ and
$n=4$ terms sufficed for a good fit, though some galaxies required up to
$n=6$.  As is well-known for bright ellipticals, the amplitude of the
$\cos(4E)$ term was the strongest of all the harmonics, which
corresponds to boxy or disky isophotes.

We note that in their SBF measurement procedure,
\citet{1990AJ....100.1416T} fit and remove large-scale residuals on
spatial scales several times the size of the PSF after subtracting the
galaxy model. The purpose of this step is to produce a very flat
background, which is needed for identification of faint objects.  These
low spatial frequencies are therefore contaminated and excluded in their
measurement of the SBF power spectra.  We chose not do this step. The
reduction of our IR data involved sky-subtraction, which is not the case
for reduction of optical imaging data. Therefore, the background of our
images is already very flat.  When cataloguing the faint sources around
the galaxy (\S~\ref{sec:sextractor}), the SExtractor software did fit
the large-scale background in the images before identifying objects, but
we did not use this fit for any other purpose.  At any rate, we did not
use such large spatial scales in fitting the galaxy power spectrum
(\S~\ref{sec:pspecfit}).

\begin{figure}
\vskip -1.7in
\vbox{\hskip -0.5in
\includegraphics[width=5in,angle=90]{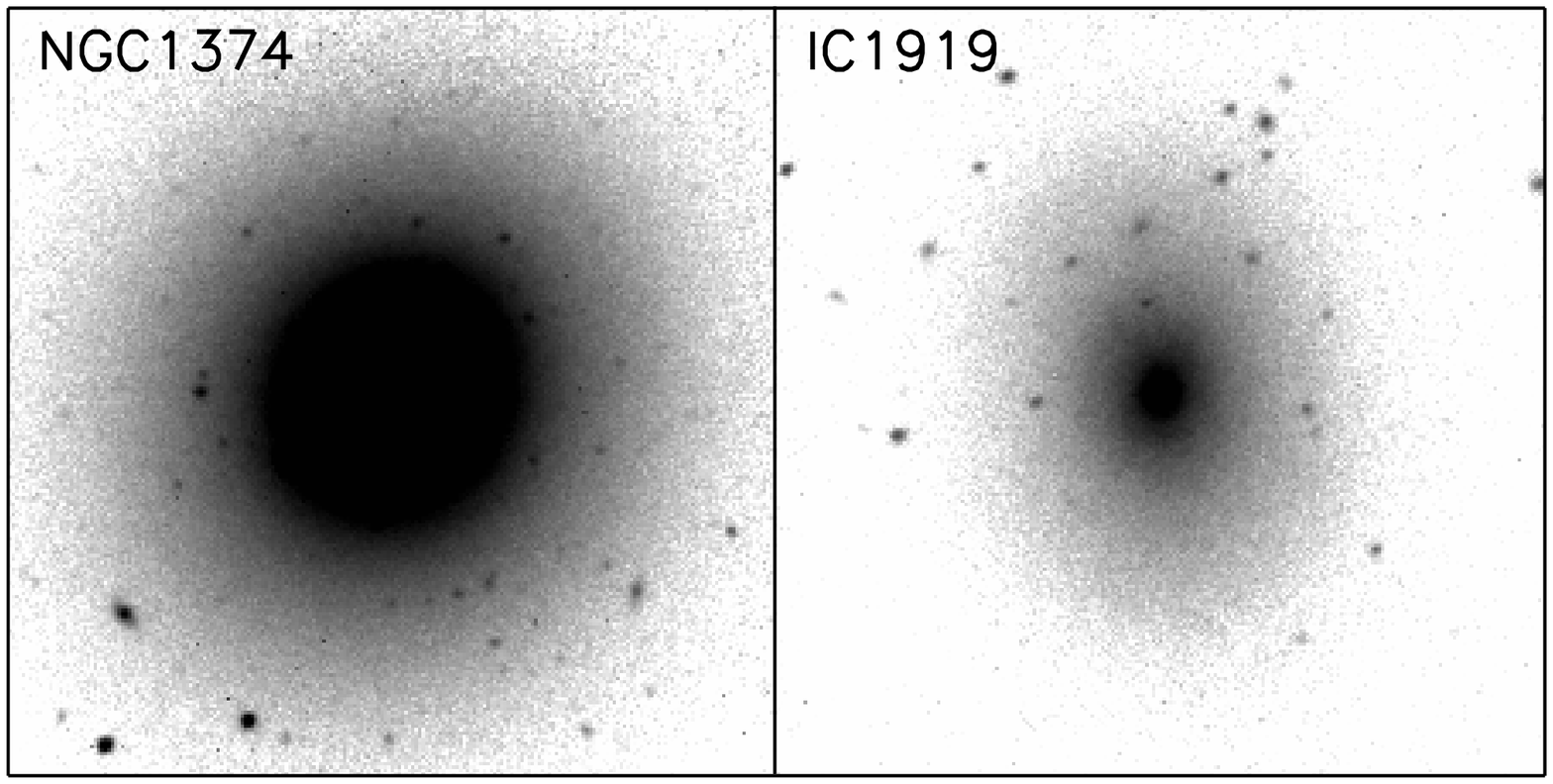}
\vskip -1.7in
\hskip -0.5in
\includegraphics[width=5in,angle=90]{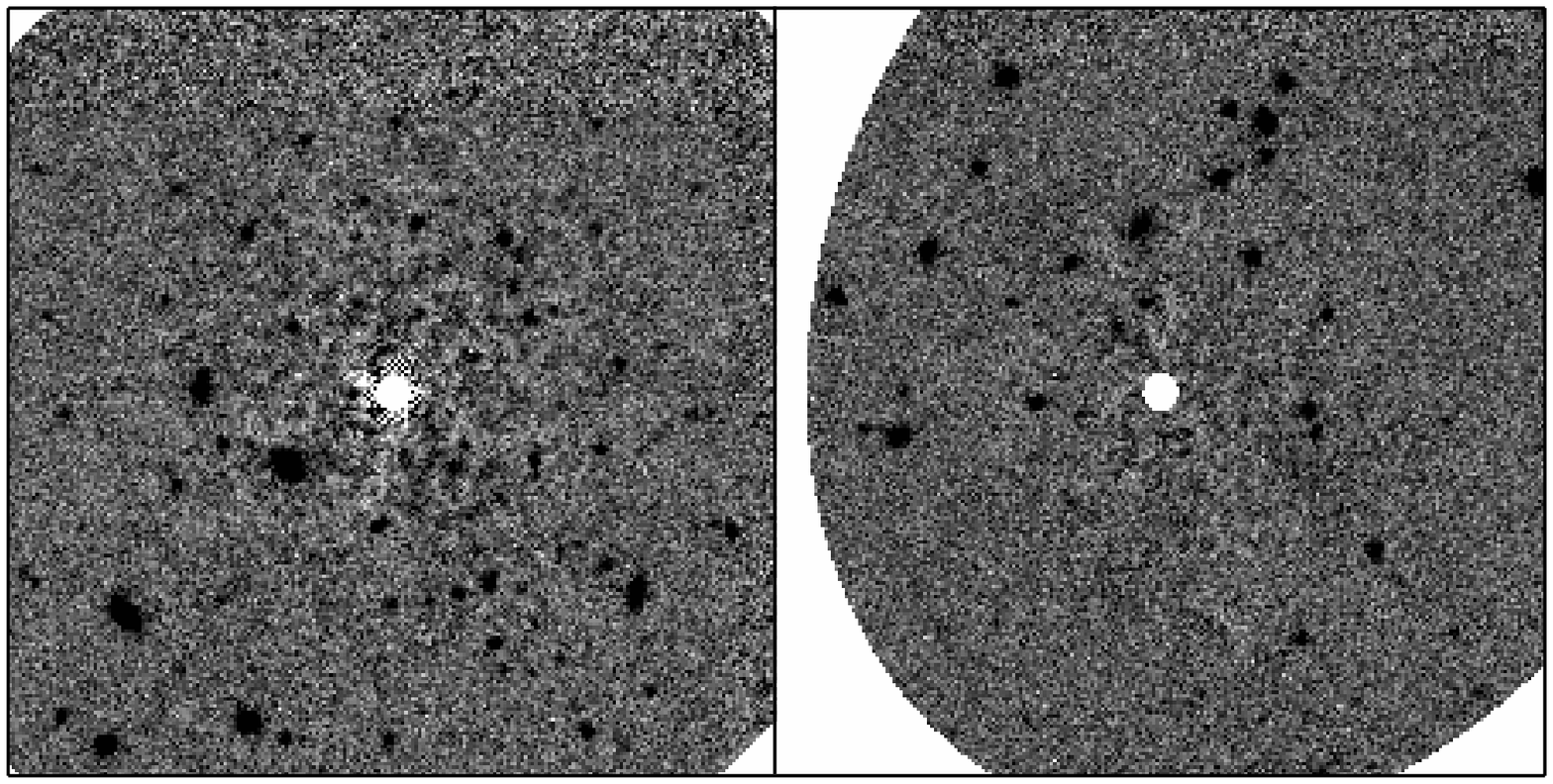}}
\caption{\normalsize {\em Top:} \Ks-band images of two Fornax galaxies,
each 1.5\arcmin\ on a side. {\em Bottom:} Same images after subtracting
an elliptical model.  The innermost region has been masked. SBFs are
seen as the faint mottling near the center, with a larger amplitude for
higher surface brightness regions. Globular clusters and background
galaxies are also present.
\label{fig-galaxies}}
\end{figure}

Figure~\ref{fig-galaxies} shows \Ks-band images for two of the galaxies
in our sample.  After subtracting the fitted galaxy model, the SBFs
become apparent, along with the contaminating globular clusters and
background galaxies.

\subsection{Measuring Globular Clusters and Background Galaxies} 
\label{sec:sextractor}

\subsubsection{Photometric Catalogs}

After subtracting the fitted model from the galaxy image, we compiled a
photometric catalog of all astronomical objects using the SExtractor
software of \citet{1996A&AS..117..393B}, version 2.0.21.  We multiplied
each pixel by the square root of its exposure time to create a mosaic
with uniform noise over the entire field.  Objects in the
noise-normalized mosaic were then identified as any set of contiguous
pixels with a total S/N~$\ge4$ within an area equivalent to a
FWHM-diameter circular aperture.  Aperture photometry was then done on
the original \Ks-band image.  We used the resulting ``${\tt
MAG\_BEST}$'' magnitudes, which for uncrowded objects use apertures
determined from the moments of each object's light distribution.  This
method is similar to that of \citet{1980ApJS...43..305K} and is designed
to recover most of an object's flux while keeping the errors low.

We ran Monte Carlo simulations to quantify the completeness of the
resulting object catalogs. These were also done to verify the accuracy
of the magnitudes measured by SExtractor.  We inserted 30 artificial
stars of known magnitude into the images and then processed the images
with SExtractor in the same fashion as the original data.  For the
artificial stars, we used the same stars as those used for the SBF power
spectra measurements, described below.  For each magnitude bin, we ran
the simulation 30~times for a total of 900~stars. We compared the
recovered magnitudes to the input ones in order to determine the random
and systematic photometric errors. The artificial stars were chosen to
span a relevant range in magnitudes in 0.25~mag steps.  Typical 50\%
completeness limits for the galaxy images were $K\approx19.5$~mag.

\subsubsection{Measuring the Luminosity Functions} \label{sec:lffit}

Globular clusters and background galaxies fainter than the detection
limit appear as unresolved point sources, which contribute to the
fluctuations. We quantified their contribution using simple analytic
approximations of their luminosity functions.
Background galaxy counts are well described by a power-law distribution
\begin{equation}
n_{gal}(m) = A \ 10 ^ {{\gamma}m}
\label{eqn:gals}
\end{equation}
where $n(m)$ is the number of galaxies per unit area per magnitude.  We
used the galaxy counts measured by \citet{1999A&A...349..751S} as these
offer the best combination of depth ($K\approx17-22$) and area coverage
(20~arcmin$^2$).  A weighted fit to their tabulated counts gives
$\gamma=0.40$ with \hbox{$A = 10^{4.73}$~deg\pertwo\ mag\perone} for
$K=21$, or one galaxy~\hbox{arcmin\pertwo~mag\perone} at $K=18.05$.  The
slope and amplitude of these counts agree well with deeper ($K\le24$)
smaller-field counts of \citet{1998ApJ...505...50B}. The slope of the
counts is somewhat larger than that from the deep imaging of
\citet{1995ApJ...438L..13D}, but the amplitude agrees reasonably well
provided that the magnitude corrections advocated by Bershady \etal\ are
used.\footnote{Our adopted parameters for the faint galaxy population
differed from \citet{1998ApJ...505..111J}, who used galaxy counts from
\citet{1994ApJ...434..114C}.  It can be shown (from equation 9 of
\citealp{1995ApJ...442..579B}) that we consequently derive a residual
galaxy variance $10^{0.1(m_{cut}-18.5)}$ larger, where $m_{cut}$ is the
cutoff magnitude. For $m_{cut}=19.5$, this factor is 1.25.}

The globular cluster luminosity function (GCLF) is well-described by a
Gaussian function
\begin{equation}
n_{glob}(m) = {{N_0}\over{\sigma \sqrt{2\pi}}} \
e^{-(m-m^0)^2 / 2\sigma^2}
\label{eqn:globs}
\end{equation}
where $N_0$ is the total number of globulars, $\sigma$ is the width of
GCLF in magnitudes, and $m^0$ is the peak (or turnover) magnitude.  To
accurately fit all three free parameters requires a large sample of
globular clusters reaching to well past the turnover magnitude.  Our
$K$-band data contained relatively few globular clusters because of the
bright sky background and the intrinsically blue color of metal-poor
star clusters. The resulting $K$-band LFs reached only a few magnitudes
above the peak of the GCLF so any fits for all three parameters would
have been poorly constrained.  Therefore, like previous optical and
near-IR SBF studies, we chose values for $m^0$ and $\sigma$ and fitted
only for the amplitude of the GCLF.

Four of the galaxies in our sample have high quality $V$-band data which
reach fainter than the GCLF peak: NGC~1344, 1380, 1399, and 1404 (see
compilation in \citealt{2000ApJS..128..431F}).  For these, we used the
measured $\sigma$ and location of the GCLF peak, converting $V$-band
$m^0$ into $K$-band by assuming $V-K=2.28$, the average of the Milky Way
and M31 globular clusters \citep{2000AJ....119..727B}.  For the
remaining galaxies, we assumed $\sigma=1.4$~mag, which is a good match
for most giant ellipticals \citep{1997eds..proc..254W,
1997AJ....114..482B,har00}.  Some of these galaxies do have GCLF
measurements, but they do not reach well past the GCLF turnover; hence,
they were not well-suited for $\sigma$ determinations
\citep[e.g.][]{1996A&A...309L..39K}.  For the GCLF turnover magnitude,
we used relative $I$-band SBF distances from \citet{sbf4} to tie the
galaxies with \HST\ GCLF data to those without, incorporating the errors
in the $I$-SBF distances into the LF fits. From the four Fornax galaxies
with high-quality $V$-band data, we found a weighted average of
$<M^0_K>=-10.04\pm0.10$~mag.  For the other galaxies, we used their
$I$-band SBF distances to compute the expected $m^0_K$.  Note that we
were in fact using the {\em relative} $I$-band SBF distances to
determine $m^0_K$; thus, we are not directly dependent on the choice of
the $I$-band SBF zero point \citep[e.g., ][]{2000ApJ...530..625T}.  Also,
as opposed to using/including Virgo cluster measurements
\citep[e.g.][]{1993AJ....105.1358S}, the use of Fornax galaxies alone
circumvents concerns about differences in $M^0_V$ between Fornax and
Virgo \citep{2000ApJ...529..745F}, perhaps due to environmental effects
\citep{1996ApJ...465L..19B}.

\subsection{Fitting the Galaxy Power Spectra} 
\label{sec:pspecfit}

We first selected a region of the galaxy to analyze.  We constructed a
software mask defining an annular region and excluding contaminating
sources (globular clusters and background galaxies).  The
model-subtracted galaxy image is then multiplied by this mask, and the
Fourier power spectrum of the central 256$\times256$ pixel region was
determined.  For most galaxies, we used an annulus spanning
4\arcsec--24\arcsec\ in radius. For the faintest galaxies (NGC~1336,
1373, 1375, 1380B, and 1419), we used a smaller annulus of
4\arcsec--16\arcsec.  Some of the galaxies had strong residuals in the
their inner regions (\eg, small disks) which were not well-fitted by
elliptical isophotes; for these, we chose a larger inner radius and
increased the outer radius to maintain the same analysis area.

At this stage, the image was composed of fluctuations due to the
galaxy's stellar population (\ie, SBFs), fluctuations from the
unresolved contaminating sources which are too faint to be catalogued,
and Poisson shot noise from the sky background.  Read noise was
negligible since the IR sky background was very bright.  In the absence
of seeing, all the pixels in the image would be uncorrelated. Therefore,
the power spectrum of the image would be white noise, \ie, equal power
at all wavenumbers.  The amplitude of the power spectrum would just be
the total variance of fluctuations and sky noise.

However, because the fluctuations are astronomical sources, they are
convolved with the point spread function (PSF) before reaching the
detector; this introduces a small-scale spatial correlation.  In
contrast, the Poisson sky noise is uncorrelated since it is not
convolved with the PSF.  Therefore, the power spectrum of the galaxy
image has two components: (1) rising power at low wavenumbers due to the
PSF-convolved fluctuations and (2) white noise due to the shot noise.
Notice that atmospheric seeing has allowed us to easily distinguish
between these two components in the power spectrum. The galaxy power
spectrum $P(k)$ can then be represented as
\begin{equation}
P(k) = P_0 \times E(k) + P_1,
\label{eqn:pspec}
\end{equation}
where $P_0$ is total variance due to the fluctuations, $P_1$ is the
white noise component, and $E(k)$ is referred to as the expectation
power spectrum. As we explain below, this is very nearly the power
spectrum of the PSF.

To measure the amplitude of $P_0$, we first measured the PSF power
spectrum.  We used the star/galaxy classifications produced by
SExtractor to identify bright PSFs in the galaxy images. A box of
25\arcsec\ was used to extract the PSF stars, and other stars within
this region were removed.  We then normalized the PSF to unity flux.
For 9 out of 19 galaxies, multiple PSFs were available, and we used the
weighted average of the resulting SBF magnitudes for the final result.
Comparing SBF measurements for the galaxies with multiple PSFs led to
an error estimate of 0.08~mag from PSF uncertainties, which included
errors in the PSF normalization.  For galaxies with only a single PSF,
we added this amount of error in quadrature.

To properly model $P(k)$, we needed to account for two additional
factors aside from the PSF power spectrum: (1) the radial variation in
the amplitude of the SBF variance, and (2) the effect of the software
mask on $P(k)$. First, in the galaxy image, the SBF variance depended on
radius since the variance scales linearly with the galaxy surface
brightness.  To account for this we created a ``window function,'' which
was the mask times the square-root of the galaxy model.  The window
function was proportional to the rms fluctuations from SBFs in the
masked galaxy image.  Next, we needed to properly account for the effect
of masking the image on the galaxy power spectrum.  It can be shown
\citep{liuthesis} that the appropriate way to do this is to convolve the
power spectrum of the window function (which contains the mask) with the
power spectrum of the PSF. This convolution is then $E(k)$, the
expectation power spectrum. The net result is that the PSF power
spectrum is slightly broadened by the mask. Once we had $E(k)$, we
simply fitted $P(k)$ to solve for $P_0$ and $P_1$ in
Eqn.~\ref{eqn:pspec}.  

The very lowest wavenumbers were not suitable for SBF analysis:
flat-fielding and sky-subtraction errors produce extra power on the
largest spatial scales which contaminate the SBF signal.  To assess
which wavenumbers to exclude, we examined the power spectra of the
reduced blank sky fields.  As described in \S~\ref{sec:redux}, for each
galaxy, observations of the sky field images were interlaced with the
galaxy images and used the same dither pattern.  Also, the sky fields
were reduced in a nearly identical fashion.  The principal difference
was that the frames used to construct each local sky frame for sky
subtraction were on average farther separated in time than those used
for sky-subtraction of the galaxy frames (this of course is inevitable).
We masked the point sources and galaxies in the reduced mosaics of the
sky fields and then examined the power spectra. We found that the
unusable wavenumbers were $k\lesssim16$, corresponding to spatial scales
larger than about 1/16 of the detector size.  Thus in fitting the galaxy
power spectra, we used only wavenumbers from $k=20$ to $k=128$, the
Nyquist frequency.

\begin{figure}
\centering
\vskip -1.2in
\hbox{\hskip -0.4in
\includegraphics[width=7in]{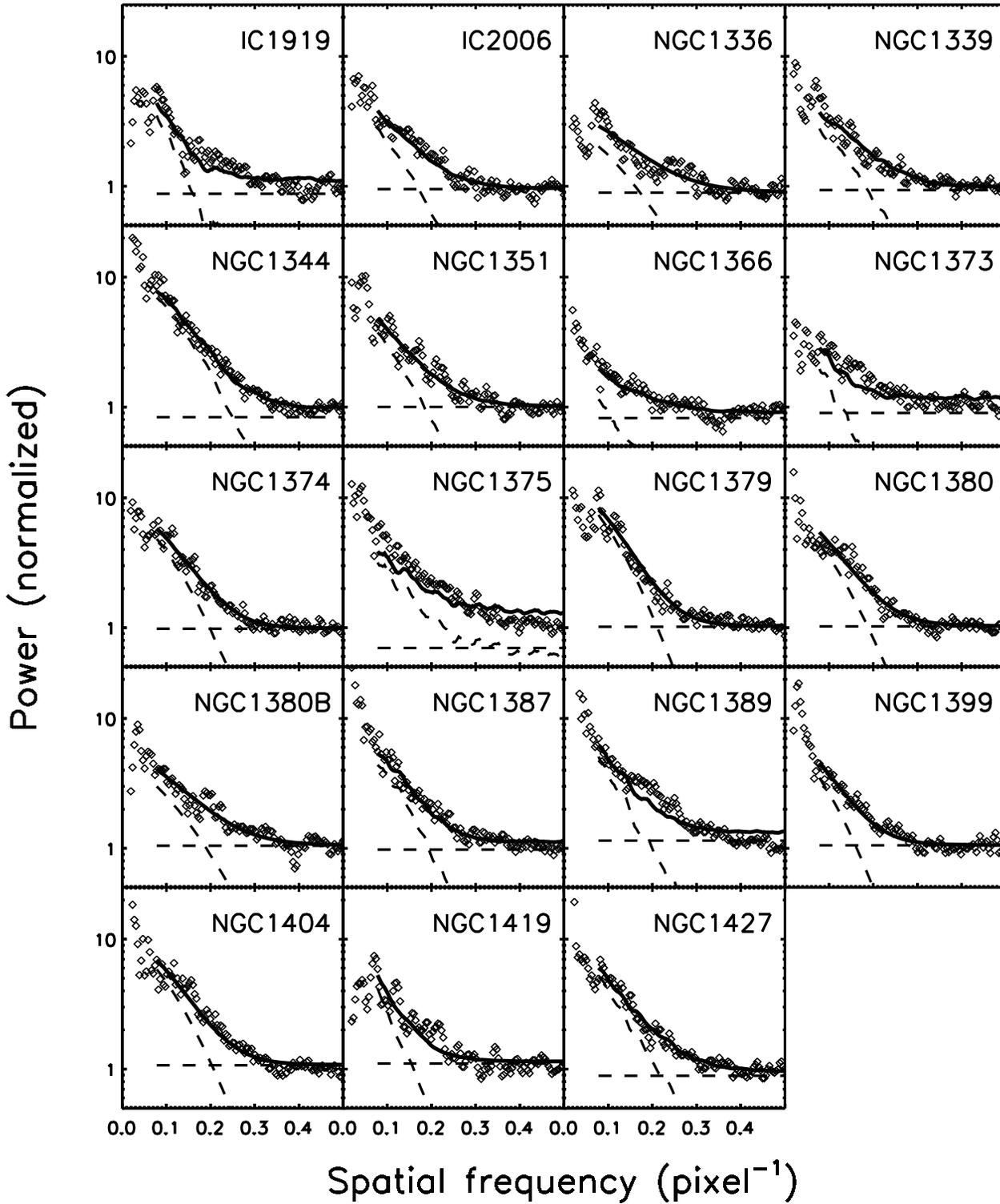}}
\vskip -0.8in
\caption{\normalsize \Ks-band fluctuation power spectra for the central
256$\times$256~pixel region of our Fornax galaxy images.  The galaxy
power spectra are fitted by the sum (solid line) of a scaled version of
the PSF ($P_0 \times E(k)$) and a constant ($P_1$); the dashed lines
show the contributions from these two components.  We use wavenumbers
from $k=20$ (spatial frequencies of 0.05~pixel\perone) to $k=128$ (the
Nyquist frequency) and do the fit with the two-dimensional power
spectra. The one-dimensional azimuthal averages are
plotted. \label{pspectra}}
\end{figure}

Figure~\ref{pspectra} shows the \Ks-band fluctuation power spectra for
our 19 Fornax galaxies and the resulting fits.  The power spectra were
two-dimensional images, and we did the fit with the full images.  To
show the results, we have plotted the azimuthally averaged
one-dimensional power spectra.

In order to gauge the errors in measuring $P_0$, we analyzed the power
spectrum of each quadrant of the galaxy image independently. We then
used the average of these four fits to obtain $P_0$, using the number of
unmasked pixels in each quadrant as the relative weights and the
standard error of the fits as the uncertainty. We ran a series of Monte
Carlo tests to determine if these quadrant-derived errors were
reasonable. Using the surface brightness profile measured for target
galaxies, we created 100 images of artificial galaxies, added SBFs with
known amplitude, convolved with a Gaussian PSF, and added Poisson shot
noise appropriate for our observations. We then fitted the resulting
fluctuation power spectra. Over a range of S/N appropriate for our data,
we found errors from this quadrant-averaging were accurate.  

For NGC 1389, the quality of the power spectrum fit appears to be
somewhat poor.  The S/N in the PSF used for the fit was good, so
potentially the problem arises from a mismatched PSF. However, with only
one PSF available in the field, we cannot directly determine if this was
the origin of the problem.  An independent observation of this galaxy
would be useful.

For NGC~1373 and NGC~1375, the PSFs used for the power spectra analysis
were somewhat lower S/N than the rest of the sample.  To gauge if the
PSFs were sufficient to accurately measure \mbar, we used the
aforementioned Monte Carlo tests to simulate the effect of noisy PSFs.
We found for the PSF and galaxy power spectra S/N of these two cases,
the errors in the \mbar\ results are expected to be negligible.  For
NGC~1380B, no suitable PSF was present in the galaxy images. We chose to
use a bright star in the sky field for this galaxy as the PSF; this had
a comparable FWHM to faint point sources in the galaxy image. (The
above-described Monte Carlo simulations also demonstrated that
systematic errors in \mbar\ measurements scale roughly linearly with the
mismatches in PSF FWHM.)

\subsection{Computing the Stellar SBF Variance}
\label{varsub}

The variance measured from the power spectrum is the sum of the variance
from the galaxy's stars, which is the signal we desire, along with the
variance from unresolved background galaxies and globular clusters.
Having fitted the LF for these contaminating sources
(\S~\ref{sec:lffit}) using analytic functions, we integrated these
functions to compute the variance below the cutoff magnitude.

For a power-law galaxy luminosity function, \citet{1995ApJ...442..579B}
show the variance per pixel due to the unmasked galaxies below the
cutoff magnitude is
\begin{equation}
\sigma^2_{gal} = \frac{p^2}{(0.8-\gamma)~{\ln10}}\ 
                 10^{0.8(m_1^{\star} - m_{cut}) - \gamma(m_g-m_{cut})}
\end{equation}
where $p$ is the pixel scale in arcsec/pixel, $m_1^{\star}$ is the zero
point of the image (magnitude of a star which produces 1 DN for the
integration time), $m_{cut}$ is the cutoff magnitude, and $m_g$ is the
magnitude where the galaxy surface density is 1
mag\perone~arcsec\perone.  All the parameters in the equation were fixed
except for $m_{cut}$, which we determined from the completeness
experiments (\S~\ref{sec:sextractor}).

Similarly, having fitted for the normalization $N_0$ of the GCLF, we
computed the variance per pixel due to the globular clusters below the
detection limit.  \citet{1995ApJ...442..579B} show that this is
\begin{eqnarray}
\sigma^2_{GC} & = & \frac{N_0}{2}\ 
  10^{0.8(m_1^{\star} - m^0_K + 0.4\sigma^2{\ln10})} \nonumber \\
              &   &  
  \hbox{erfc}\left(\frac{m_c-m^0_K+0.8\sigma^2\ln10}{\sqrt{2}\sigma}\right)
\end{eqnarray}
where $m_1^{\star}$ is the zero point of the image, $m^0_K$ is the
$K$-band GCLF peak magnitude, $\sigma$ is the GCLF width in magnitudes,
and erfc($x$) is the complementary error function.  When calculating
errors for $\sigma_{GC}^2$, we included the formal fitting errors on
$N_0$ as well as the errors in the $I$-band SBF distances to the
galaxies which governs the value of $m^0_K$ used in the GCLF fitting.

The fitted amplitude of the power spectrum, $P_0$
(eqn.~\ref{eqn:pspec}), was then corrected for the variance $P_r$ due to
the unresolved globular clusters and galaxies.  The quantity $P_r$ was
the sum of the galaxy and globular cluster variances ($\sigma_{gal}^2$
and $\sigma_{GC}^2$, respectively) divided by the mean galaxy surface
brightness per pixel in the fitting region. This contaminating variance
amounted to about 10--30\% of the total variance. The remaining quantity
($P_0-P_r$) was then the variance due solely to the stellar SBFs.  This
was converted to an apparent magnitude, $\mbar_{K_S}$, using the
measured photometric zero point and extinction correction.  The final
errors on \mbarKs\ comprised the quadrature sum of the errors in the
photometric calibration (0.02~mag), the PSF uncertainty for galaxies
with only a single PSF (0.08~mag), and the measurement errors in $P_r$
and $P_0$.


\section{Results} \label{sec:results}

Table~\ref{table-sbfmags} summarizes our measurements. For each galaxy,
we tabulate the mean \Ks-band surface brightness in the measurement
region, the ratio of $(P_0\!-\!P_r)/P_1$ as a measure of the S/N of our
data, the final \Ks-band SBF apparent magnitudes (\mbarKs), the \Ks-band
SBF absolute magnitudes (\MbarKs) derived from $I$-band SBF distances to
individual galaxies and a Cepheid distance to the Fornax cluster, and
the (distance-independent) \Ibar--\Ksbar\ SBF color.

We have five galaxies in common with the \Kp-band (1.9--2.3~\micron)
measurements of \citet{1998ApJ...505..111J}. The overall agreement
between our \mbar's is good, with an average difference of
0.12$\pm$0.04~mag (0.07$\pm$0.06~mag if we exclude their low S/N data
for NGC~1339), in the sense that our measurements tend to be fainter.
Our data use the \Ks-band (2.0--2.3~\micron) filter, and the difference
in filter is expected to make the Jensen \etal\ data about
0.01--0.02~mag fainter according to theoretical models
\citep{bc2000sbf}. Still, our measurements of the fluctuation apparent
magnitudes tend to be slightly fainter, in part because of our different
adopted parameters for the background galaxy LF (\S~\ref{sec:lffit}).

To compute the \Ks-band SBF absolute magnitudes and SBF colors, we used
\Ibar\ data and $I$-band distances from \citet{sbf4}. The latter are
calibrated using \HST\ Cepheid distances from
\citet{2000ApJS..128..431F} for six nearby spiral galaxies which also
have $I$-SBF measurements.  The tabulated errors for the absolute
\Ks-band SBF magnitudes include the errors from the $I$-SBF distances
added in quadrature.  However, the listed errors {\em do not} include
the systematic error in the \HST\ Cepheid distance scale, believed to be
about $\pm$0.16~mag \citep{2000ApJ...529..786M}, which is mostly due to
uncertainty in the distance to the Large Magellanic Cloud.

We also computed \MbarKs's by using an \HST\ Cepheid distance to the
Fornax cluster. We adopted a Cepheid distance modulus of $31.42 \pm
0.06$~mag (19.2~Mpc) from the weighted average of distances to NGC~1326A
and NGC~1365 \citep{2000ApJS..128..431F}; like
\citet{2000ApJ...530..625T}, we excluded NGC~1425 as part of Fornax.
(We discuss the effect of the new Cepheid distances from \citealp{fre01}
in \S~\ref{sec:Mbarcalibrate}.) We also added in quadrature a random
error to account for the cluster's depth along the line of sight
(0.06~mag). Just like the $I$-SBF calibrated $\Mbar_{K_S}$'s, we did not
include the systematic error in the \HST\ Cepheid distance scale. Note
that the total errors for the Cepheid-calibrated \MbarKs\ are smaller
than the $I$-band SBF calibrated ones. This is because the error added
in quadrature from the Cepheid distance was only 0.08~mag, whereas the
median error for the $I$-SBF distances to individual galaxies was
0.21~mag.

If the Cepheid cluster distances are used, we are assuming the spiral
galaxies used for the Cepheid measurements and the elliptical/lenticular
galaxies with SBF data lie at the same distance.  However, there is
evidence that spirals in groups tend to lie at larger radii, with the
ellipticals in the inner regions \citep{1992PASP..104..599J,
2000ApJ...529..768K}.  We circumvented this problem by choosing to use
$I$-band SBF distances to individual galaxies. Although the $I$-SBF
calibration is based on spirals with both \HST\ Cepheid and $I$-SBF
measurements, there are still some lingering concerns: SBF measurements
in spiral bulges are challenging, and bulges and ellipticals might have
different stellar populations \citep[e.g.,][]{1997ARA&A..35..637W}.


\section{Analysis}
\label{sec:analysis}

Our new \Ks-band sample of early-type galaxies in Fornax has more than
doubled the number of high-quality IR SBF measurements.  The sample
covers an expanded range of galaxy properties (\eg, color, velocity
dispersion, luminosity, and morphological type) compared to past
studies. In particular, our new measurements include galaxies which are
fainter and have bluer integrated colors.  Also, all previous
high-quality $K$-band SBF measurements were of ellipticals (except for
the bulge of M31); we have added several S0 galaxies.

In the analysis which follows, we combine our Fornax sample with 10
galaxies in nearby clusters from the literature.  We use data for M32
and the bulge of M31 from \citet{1993ApJ...410...81L}.  Measurements for
one Eridanus cluster (NGC~1407) and five Virgo cluster galaxies
(NGC~4365, 4406, 4472, 4552, 4636) come from
\citet{1998ApJ...505..111J}.  We do not include the lower S/N data of
Jensen \etal\ given the concerns about systematic biases in SBF
measurements at low S/N \citep[see][]{1996ApJ...468..519J}. For the same
reason, we do not include data from \citet{1994ApJ...433..567P}, and
also because these authors did not correct their SBF measurements for
globular cluster contamination.  Likewise, we hold our own Fornax data
to similar standards and exclude \mbarKs\ measurements with S/N$<$3,
namely those for IC~1919, NGC~1366, and NGC~1373.

These literature data were all taken with the \Kp\ filter
(1.9--2.3~\micron; \citealp{1992AJ....103..332W}).  To compare these to
the slightly redder bandpass of our \Ks\ filter, we subtracted 0.02~mag
from the published data as suggested by the SBF models of
\citet{bc2000sbf}.  We also changed the extinction corrections from the
\ion{H}{1}-derived values of \citet{1984ApJS...54...33B} to the
DIRBE/IRAS-derived values of \citet{1998ApJ...500..525S} and used the
latest \VmI\ galaxy colors from \citet{sbf4}.

Finally, we include \Ks-band data for NGC~3379 in Leo and NGC~4489 in
Virgo from \citet{mei2001}. 


\subsection{Calibration of \MbarKs}
\label{sec:Mbarcalibrate}

\subsubsection{Calculations}

Table~\ref{table-calibrate} summarizes our various calibrations for
\MbarKs. There are several choices. We provide calibrations based on
only our Fornax data or the total sample. Using only our Fornax data
ensures the calibrations are based on measurements (and their associated
errors) determined in a homogeneous fashion.  For distances to galaxies,
we use either individual distances to galaxies from $I$-band SBF
\citep{sbf4} or Cepheid distances to the galaxy clusters as a whole.  We
adopt Cepheid distance moduli of $24.44\pm0.10$~mag for M31 and M32,
$30.08\pm0.06$~mag for Leo, and $31.03\pm0.06$~mag for Virgo
(\citealp{2000ApJ...529..745F}; see below for a discussion of the new
distances by \citealp{fre01}).

In averaging the data, we either take a weighted average to derive a
universal zero point for \MbarKs, or we fit for a linear trend with \VmI\
to account for stellar population differences between galaxies, as is
done for $I$-band SBFs.  The data clearly favor a dependence on \VmI. A
non-parametric evaluation using the Spearman rank-order correlation
coefficient ($r_s$) finds a probability of only $3\times10^{-3}$ for the
dependence to occur by chance.  Also, adopting a linear relation gives a
lower reduced chi-square (\rchisq) than a universal zero point, and the
derived slopes are statistically different from zero.  Finally, the
existence of a color dependence is to be expected, given that $I$-band
SBFs demonstrate substantial stellar population variations between
galaxies.

We fit \MbarKs\ versus \VmI\ accounting for errors in both
variables. For calibrations using the Cepheid distances, we use the
algorithm described in \citet{1992nrca.book.....P}.  However, for
calibrations using the $I$-band SBF distances, a different procedure is
required. Since the $I$-band SBF distances depend on the \VmI\ color, or
more specifically $4.5\times(\VmI)$, errors in \VmI\ will directly lead
to errors in the resulting \MbarKs.  We need to take into account this
non-zero covariance when fitting for the relation between \VmI\ and
\MbarKs.  Such a covariance would bias the fitting towards a slope of
4.5, with more severe bias as the fractional errors in \VmI\ become
larger. This effect was not accounted for by \citet{1998ApJ...505..111J}
in their fits.

To avoid this bias, we start with the maximum likelihood approach
described by \citet{stet89} which handles independent errors in both
variables. We then enhance the method to account for the fact that the
error ellipses for each datum are tilted, \ie, the covariance is
non-zero.  We validate our algorithm by applying it to simulated data
sets.\footnote{We test our algorithm using simulated data sets obeying
$y = A + Bx$, where the errors in $x$ and $y$ are partially
correlated. It is straight-forward to compute the shape of the tilted
error ellipses \citep{cowan98}. When the covariance is zero, our
approach produces very similar results to the heuristic method of
\citet{1992nrca.book.....P}.  For non-zero covariance, our approach is
far more effective, producing essentially unbiased fits and accurate
error estimates for $A$ and $B$.  Our simulations span a wide range of
the parameter space in the coefficients, the sizes of the errors, and
the amount of covariance, including values comparable to the
observations.}
These numerical experiments demonstrate that if we simply use an
ordinary least-squares approach, or the method given by Press \etal, the
bias in the fits of \MbarKs\ versus \VmI\ would small but not
insignificant.  The reason why the bias is relatively small is because
of the high accuracy of the \VmI\ measurements.


For all the calibrations, we used a standard $\sigma$-clipping scheme to
avoid the undue influence of a few outliers. This resulted in the
exclusion of NGC~1389 and 1419, both of which have very bright \Ks-band
fluctuations ($\MbarKs\approx-6.6$~mag); these galaxies are discussed in
the next section.  More sophisticated iterative fitting schemes, \eg,
based on robust weighting schemes described in
\citeauthor{1992nrca.book.....P}, gave similar results. The main
advantage of the $\sigma$-clipping scheme is its conceptual simplicity.

\begin{figure}
\centering
\hbox{\hskip -0.4in
\includegraphics[width=5in,angle=90]{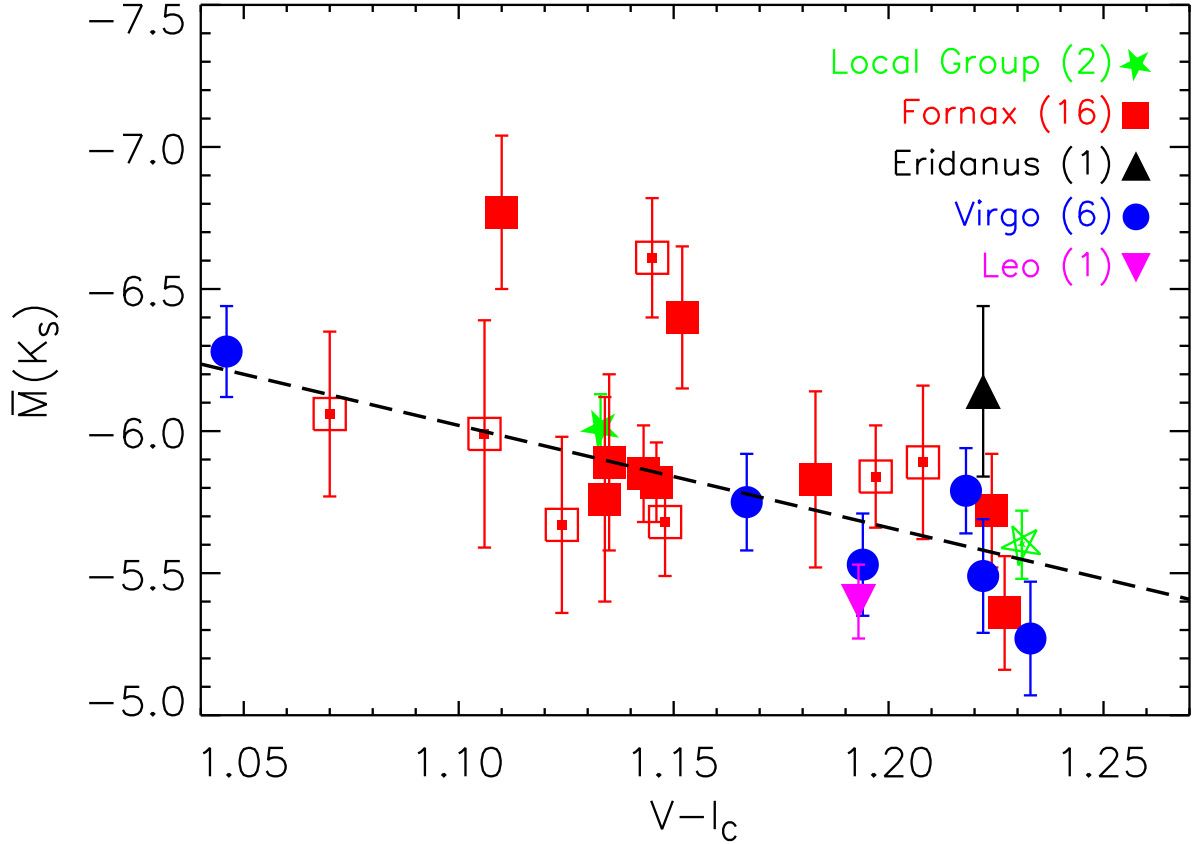}}
\caption{\normalsize Combined \MbarKs\ data set for galaxies with high-quality SBF
measurements plotted against integrated V--\Ic\ galaxy colors. Distances
to the individual galaxies come from $I$-band SBF \citep{sbf4}. The open
symbols are S0's and spiral bulges (M31), and the filled ones are
ellipticals.  The Fornax data are from this work; the Local Group (M31
and M32, with M32 being the galaxy with bluer \VmI) data are from
\citet{1993ApJ...410...81L}; and the Virgo, Eridanus, and Leo data are
from \citet{1998ApJ...505..111J} and \citet{mei2001}.  The values in
parentheses are the number of galaxies in each group with SBF
measurements. The dashed line is the best fit.  The fit excludes the two
Fornax galaxies with very bright fluctuations ($\MbarKs \approx
-6.6$~mag), NGC~1419 (E) and NGC~1389 (S0). The reddest galaxy (lower
right of plot) is NGC~4636 in Virgo.
\label{plot-combine}}
\end{figure}

Our preferred calibration relies on $I$-band SBF distances to individual
galaxies, as described in \S~\ref{sec:results}.  Using the total sample
with these distances gives
\begin{equation}
\MbarKs = (-5.84\pm0.04) + (3.6\pm0.8) [(V-\Ic)_0 - 1.15]\ .
\end{equation}
Figure~\ref{plot-combine} illustrates this calibration.  Note that
without accounting for the non-zero covariance between \VmI\ and
\MbarKs, the fitted slope would have been $4.1\pm0.8$, with little
change in the zeropoint.  (As expected, if we do not account for the
covariance, the fitted slope is biased towards a slope of 4.5.) There is
a hint in the data that the later-type galaxies may obey a different
calibration.  A fit to the 7 objects (6 S0's and the bulge of M31) finds
a slope of 1.8$\pm$1.5. However, the sample is small and the
significance is low.  A larger sample is needed to test this
possibility.

\subsubsection{Implications}

The dependence of \MbarKs\ with \VmI\ color is a new finding. A hint of
this effect was seen in the data set of \citet{1998ApJ...505..111J},
though it covered only about half the range in \VmI\ that our sample
does.  The dependence is seen clearly with the addition of Fornax
galaxies with $\VmI\lesssim1.15$.  The \MbarKs\ slope is nearly as steep
as for $I$-band SBFs ($4.5\times(\VmI)$; \citealp{2000ApJ...530..625T}),
meaning accurate optical photometry improves the precision of $K$-SBF
distances.  Moreover, the discovery of the slope eliminates published
concerns about anomalously bright $K$-band SBFs for M32
\citep{1993ApJ...410...81L} and NGC~4489 in Virgo
\citep{1994ApJ...433..567P, 1996ApJ...468..519J,mei2001}.  Both galaxies
lie on the plotted calibration. (NGC~4489 has the bluest \VmI\ color in
the sample, but if we remove it, the fitted zero point and slope of
change negligibly.)  However, we do find a few Fornax galaxies seem to
have very bright \Ks-band fluctuations, more than 0.5~mag above the
observed mean relation.  We address the origin of these in the next
section.  Most of these galaxies are bluer and much less luminous than
galaxies used for cosmological SBF distance measurements.

The sense of the observed trend, with redder galaxies having fainter
\Ks-band fluctuations, is opposite that predicted from theoretical
models based on single-burst stellar populations.  For the published SBF
models, theoretical calibrations based on ensemble averages of
single-burst models \citep[e.g.,][]{1993ApJ...409..530W} tend to produce
brighter fluctuations for redder galaxies.  The disagreement between
these theoretical calibrations and the observations is not surprising
given the relatively simply approach of averaging models with a range of
ages and metallicites.  Recently, \citet{bla00} have
generated more complicated models with three bursts of star formation.
They select the ages and metallicities of the bursts to be consistent
with the observed optical/IR SBF colors and the $I$-band SBF
slope. Their resulting models have IR SBF magnitudes which become
brighter with redder integrated colors, with a slope of 2.9 in the
$K$-band, consistent with our observations.  The zero point from their
models is about $\sim$0.5~mag fainter than the observed
calibration. (See \S~\ref{sec:starform} for more discussion about the
theoretical models.)

The discovery of a dependence of \MbarKs\ on \VmI\ color has important
implications for \Ho\ measurements using IR SBFs
\citep{1999ApJ...510...71J,jen01,liu2001}.  In particular, \citet{jen01}
have used \HST\ NICMOS $F160W$ (1.6~\micron) data to measure SBFs in a
sample of distant galaxies.  They assumed the $F160W$ SBFs were
invariant from galaxy to galaxy.  However, their set of nearby
calibrators were on average $\approx$0.05~mag bluer in \VmI\ than the
set of distant galaxies used to determine \Ho.  If the color dependence
of $F160W$ SBFs is as steep as that for \Ks-band SBFs, this means a
difference of $\sim$0.2~mag in the SBF absolute magnitudes between the
two sets.  The consequence would be that the measured distances for
their distant set are overestimated by $\sim$10\% and likewise the
resulting \Ho\ underestimated.

Calibrations of \MbarKs\ based on Cepheid group distances have higher
\rchisq's since the Cepheid-calibrated \MbarKs's possess much lower
errors than $I$-SBF calibrated \MbarKs's, as described in
\S~\ref{sec:results}.  We have already added in quadrature an error term
to account for the cluster depth (0.06~mag for Fornax and 0.08~mag for
Virgo).  To bring the Cepheid calibration of \MbarKs\ to
$\rchisq\approx1$, we would need to add in quadrature $\approx$0.15~mag
of error.  This result provides an estimate on the intrinsic cosmic
scatter in \MbarKs. (This calculation depends on the errors of \mbarKs\
being accurate; these should be confirmed by re-observing at least a
portion of the sample.)  For comparison, the estimated cosmic scatter in
$I$-band SBFs is about 0.05~mag \citep{1997ApJ...475..399T}, about a
factor of three smaller. This difference is not surprising; stellar
population models predict $I$-band SBFs will be degenerate in age and
metallicity once corrected for differences in galaxy colors while
$K$-band SBFs will vary more strongly depending on the underlying
population \citep{1993ApJ...409..530W, bc2000sbf}.

As mentioned before, this calibration does not account for any
systematic error in the current \HST\ Cepheid distance scale.  For
instance, recent comparisons of the Cepheid and maser distance
measurements to the nearby spiral NGC~4258 suggest the Cepheid distances
might be overestimated \citep{1999Natur.400..539H, 1999Natur.401..351M,
new01}. In this work, we have used the Cepheid distances from
\citet{2000ApJS..128..431F} since these are used to calibrate the
$I$-band SBF distances compiled by \citet{sbf4}.  \citet{fre01} have
presented new Cepheid distances by the \HST\ \Ho\ Key Project, which
incorporate several changes in the analysis. These include use of an
improved Cepheid period-luminosity relation, revised photometric
calibrations, and a correction for the metallicity dependence of the
Cepheids.  The sign and amplitude of the metallicity correction remains
quite uncertain.  Using the new distances without the metallicity
correction leads to the $I$-SBF zero point becoming fainter by
$\approx$0.10--0.15~mag. Our \Ks-band SBF calibration would change by
the same amount. However with the metallicity correction adopted by
Freedman \etal, the $I$-SBF zero point is only slightly changed,
becoming fainter by 0.06~mag.


\subsection{Recent Star Formation in Cluster Ellipticals}
\label{sec:starform}

\begin{figure}
\centering
\hskip -0.4in
\vbox{
\includegraphics[width=3.5in,angle=90]{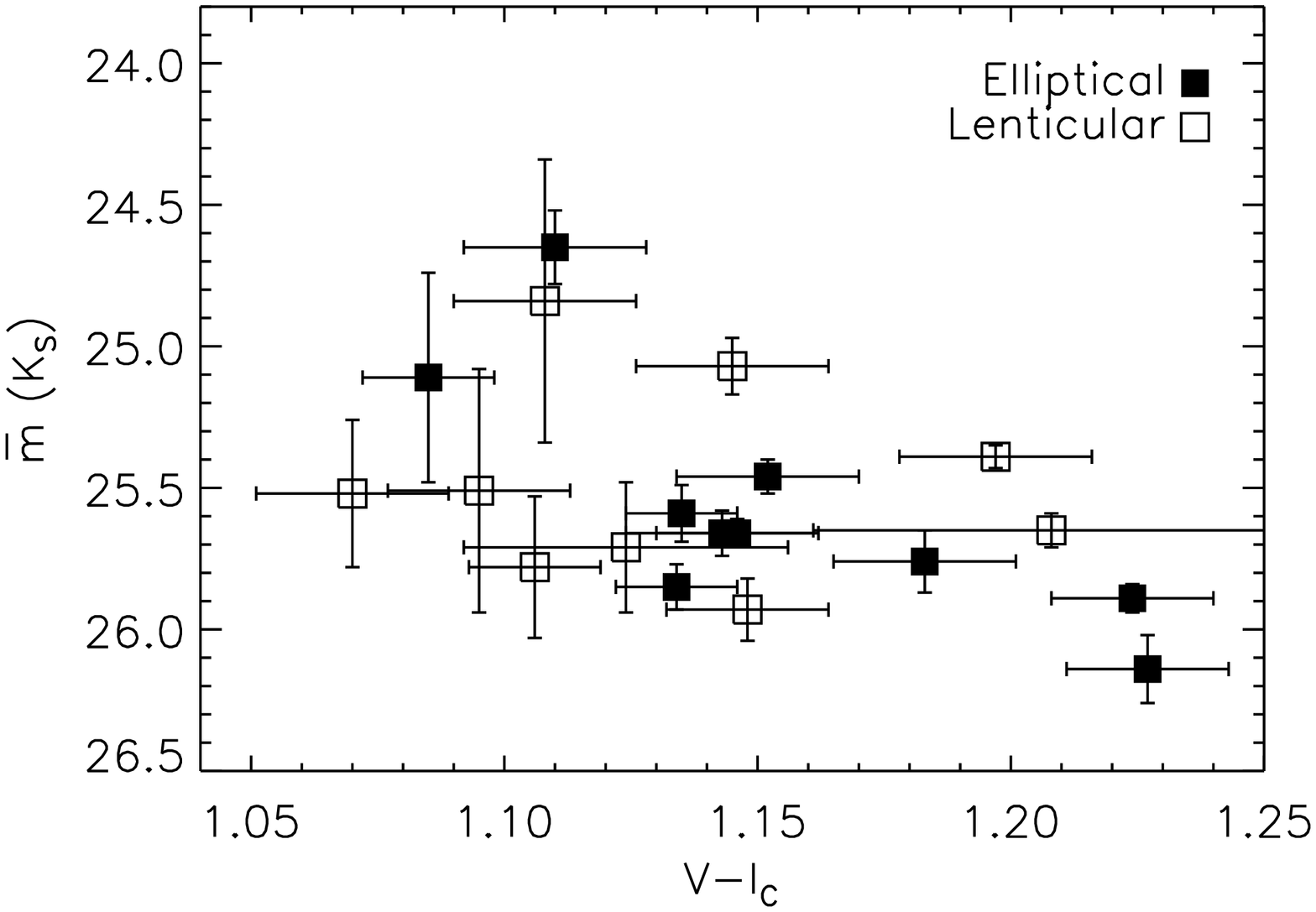}
\includegraphics[width=3.5in,angle=90]{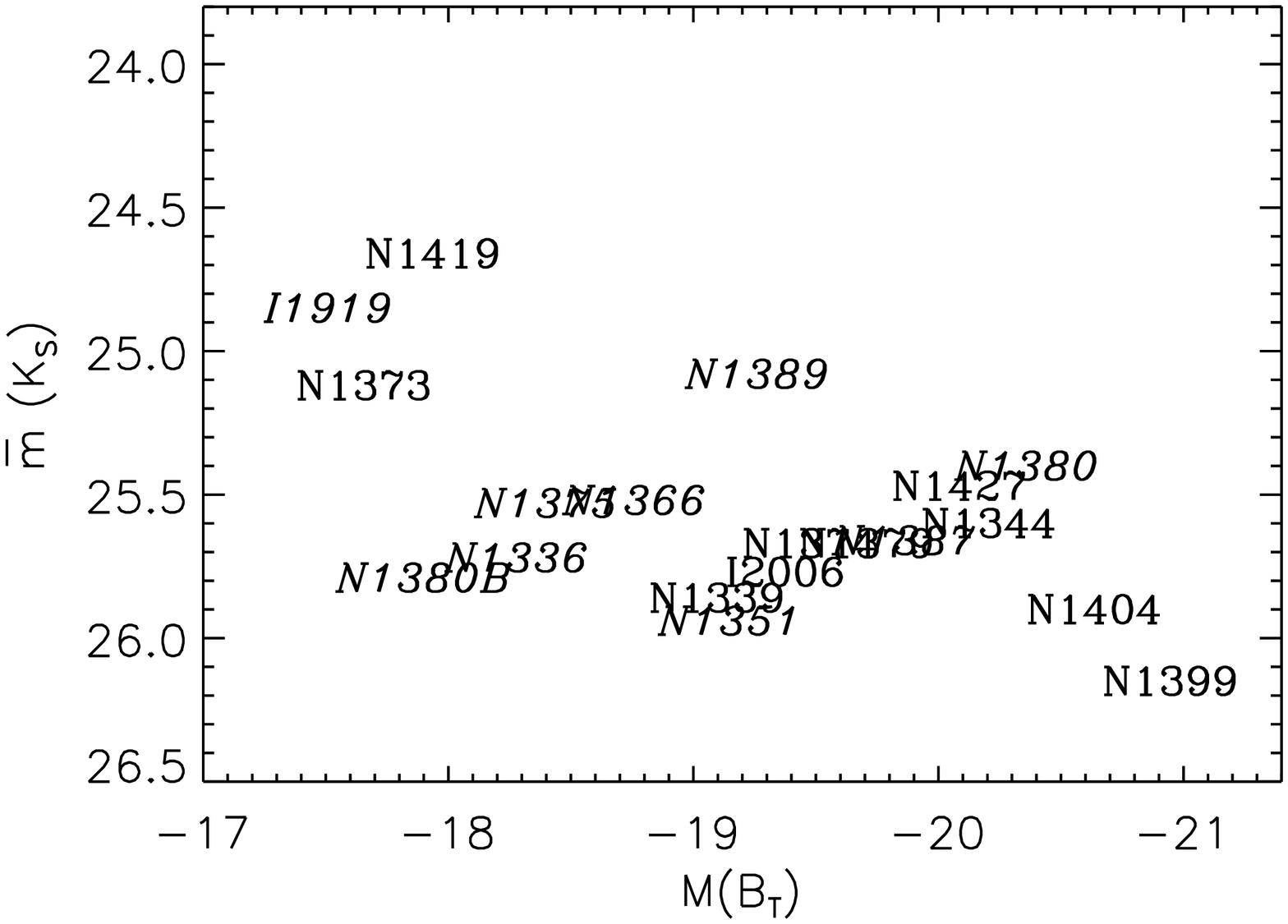}}
\caption{\normalsize \Ks-band SBF apparent magnitudes for our Fornax cluster sample
as a function of galaxy properties.  {\em Top:} \VmI\ galaxy color. {\em
Bottom: } Absolute $B$-band magnitude, with apparent magnitudes $B_{T}$
from \citet{RC3} and distances and errors from $I$-band SBF data of
\citet{sbf4}.  Elliptical galaxies are listed in roman type and S0's in
italics.  The errors in \mbarKs\ for the brighter half of the sample
($M(B_T) < -19$) are comparable to the height of the labels; for the
fainter half, the errors are a few to several times larger than the
label height.  Errors in $M(B_T)$ are comparable to or smaller than the
width of the labels.
\label{plot-fornax}}
\end{figure}

The distribution of galaxies in Figure~\ref{plot-combine} has at least
three characteristics: the elliptical and lenticular galaxies appear to
have comparable \MbarKs; the \Ks-band SBFs grow brighter as the
integrated galaxy colors become bluer; and a few galaxies have SBFs much
brighter than the bulk of the sample.  These trends are also seen in
Figure~\ref{plot-fornax} which plots the Fornax cluster apparent SBF
magnitudes, which are independent of the chosen galaxy distances, as a
function of galaxy properties.  These phenomena provide new clues to the
star formation histories of the cluster galaxies.  While a complete
analysis is deferred to a subsequent paper (M.\ Liu \etal, in
preparation), we address here some basic considerations.

Since early-type galaxies follow a color-magnitude relation (more
luminous ones have redder integrated colors
[\citealp{1977ApJ...216..214V, 1992MNRAS.254..589B}]), the correlation
between \MbarKs\ and \VmI\ color can also be seen as one between
\MbarKs\ and galaxy luminosity/mass.  Thus the bluer, less massive
galaxies tend to have brighter $K$-band SBFs.  As we discuss below,
stellar population models imply this trend reflects an age spread, with
less massive galaxies have more extended star formation histories.
Furthermore, the bluer, lower luminosity galaxies (fainter than
$M_B\approx-19$) span a wider range in \MbarKs.  This suggests the star
formation histories of these galaxies, in addition to being more
extended, are also more heterogeneous than those of the redder, more
massive galaxies.  A similar phenomenon may exist among early-type Virgo
cluster galaxies, as the lower-mass ($\sigma<100$~\kms) galaxies exhibit
much larger scatter in their Balmer and metal absorption lines than the
more massive ones \citep{con00}.

\begin{figure}
\centering 
\vskip -0.25in
\hbox{\hskip 0.75in
\includegraphics[width=5in,angle=0]{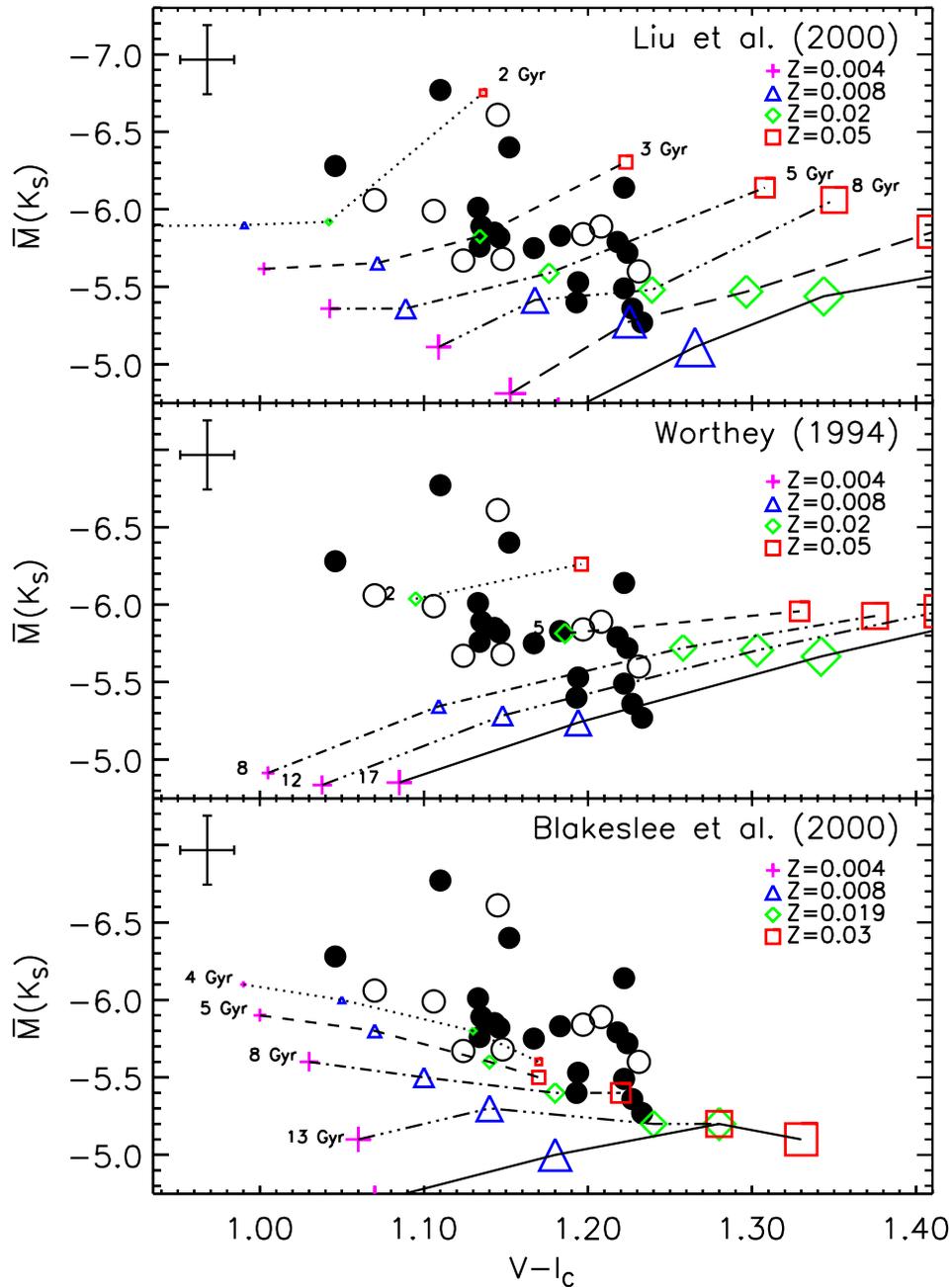}}
\caption{\normalsize Comparison of \Ks-band SBF measurements with three sets of
stellar population synthesis models. The data are the same as in
Figure~\ref{plot-combine}, using $I$-band SBF distances to individual
galaxies.  Filled circles ($\bullet$) are ellipticals, and open circles
($\circ$) are lenticular galaxies and spiral bulges. The average error
bars are plotted in the upper right corners.  Models of a fixed
metallicity have the same symbol, with increasing symbol size
representing increasing age; solar metallicity models are plotted with a
diamond ($Z=0.02$, 0.02, and 0.019). Line connect models with the same
age.  The \citet{bc2000sbf} models are 2, 3, 5, 8, 12, and 17~Gyr
old. The \citet{1994ApJS...95..107W} models cover the same, except they
are unavailable for young, metal-poor populations. The models from
\citet{bla00} are 4, 5, 8, 13, and 18~Gyr; {\em note that their
metal-rich models are plotted with the same symbols as the other models
but have different $Z$ values}.  \label{plot-models-Mbar}}
\end{figure}

Figure~\ref{plot-models-Mbar} compares the \Ks-band SBFs with three sets
of stellar population synthesis models, those of \citet{bc2000sbf},
\citet{1993ApJ...409..530W}, and \citet{bla00}.
For the first set, the models used here are a slightly revised version
of those published in Liu \etal\ --- see the Appendix for details. The
latter two models are calculated for slightly different $K$-band filters
than used for our observations, but the differences are negligible for
our purposes here. All the plotted models are single-burst stellar
populations (SSPs), where the stars form coevally and then evolve
passively. The models all use differing sets of evolutionary tracks and
stellar spectral libraries (see discussions in
\citealp{1996ApJ...457..625C}, \citealp{bc2000sbf}, and
\citealp{bla00}). As a consequence, the SSP-equivalent ages and
metallicities inferred for the galaxies depend on the choice of models.

The Liu \etal\ models indicate that the stellar populations dominating
the $K$-band SBFs of most galaxies are around solar metallicity with a
factor of 3--4 spread in age, including some as young as $\approx$2~Gyr.
Given the measurement errors in \MbarKs, the data also allow for about a
factor of two spread in metallicity. The Worthey models offer
qualitatively similar results, though with a lower mean metallicity.
Unfortunately, the Worthey models are unavailable for young, metal-poor
populations so they do not cover the full observational locus.  The
Blakeslee \etal\ models suggest a comparable spread in age and a much
wider spread in metallicity, extending above their maximum available
metallicity (1.5$\times$ solar).  These models suggest the effects of
age and metallicity are nearly degenerate for metal-rich SSPs.  It is
worth noting that despite the disagreements in the model predictions,
all three sets of models in Figure~\ref{plot-models-Mbar} indicate ages
younger than 5~Gyr for many galaxies. Such young ages imply these
galaxies have very luminous giant branches, as is seen in
intermediate-age ($\approx$0.5--5~Gyr) Magellanic Cloud clusters
\citep{1979ApJ...232..421M, 1980ApJ...239..495F, 1980ApJ...240..464M,
1981ApJ...249..481C}.

The disagreement is most severe between the Liu \etal\ and Blakeslee
\etal\ predictions, which are derived from the \citet{bc2000} and
\citet{1996ApJS..106..307V} models, respectively.  Note that the two
sets of models are computed for different metallicities; in particular,
the Blakeslee \etal\ models are not as metal-rich. At metallicities of
solar and above, which are most relevant to elliptical galaxies, the Liu
\etal\ models predict the $K$-band SBFs grow brighter with increasing
metallicity, as would be expected based on the steady increase with
metallicity in the $K$-band magnitude of the tip of the red giant branch
for Galactic globular clusters \citep{2000AJ....119.1282F}. In contrast,
Blakeslee \etal\ predict the IR SBFs do not change much between solar
and super-solar metallicity.  For the reddest ellipticals
($\VmI\gtrsim1.2$), the metallicities inferred in
Figure~\ref{plot-models-Mbar} from their models are very large.
Furthermore, for SSPs younger than $\approx$8~Gyr, Blakeslee \etal\
predict that the IR SBF magnitudes {\em decrease} with increasing
metallicity. This ``inversion'' is surprising given two known aspects of
AGB stars, which are expected to contribute a substantial fraction of
the IR SBF signal \citep[see][]{bc2000sbf}.  First, the main-sequence
turnoff mass increases with metallicity at fixed age
\citep{1988ARA&A..26..199R}.  Since AGB stars follow a core
mass-luminosity relation, higher metallicity is expected to lead to
brighter AGB stars.  Likewise, solar-metallicity intermediate-age
clusters are predicted to have AGB stars with higher luminosities than
lower metallicity clusters (\citealp[see][]{1998AJ....116...85S} and
\citealp{bc2000sbf}). Both of these argue for brighter AGB stars with
increasing metallicity, hence brighter infrared \Mbar's.
For these reasons, we favor the Liu \etal\ models, though the issue
remains to be resolved.

\begin{figure}
\vskip -0.5in
\centering
\hbox{\hskip 0.75in
\includegraphics[width=5in,angle=0]{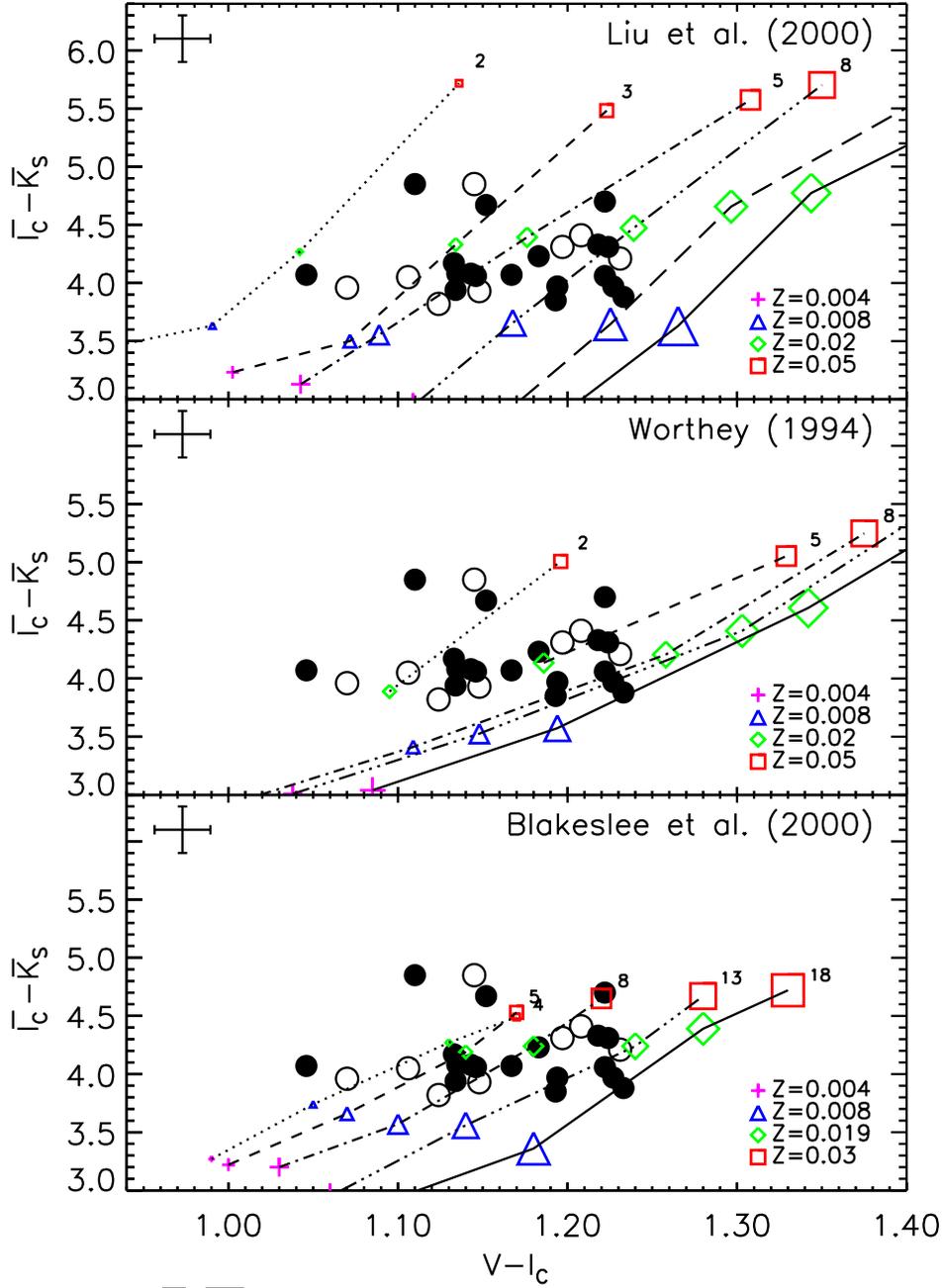}}
\caption{\normalsize Comparison of \Ibar$-$\Ksbar\ SBF colors with stellar
population synthesis models. The data and models are the same as in
Figure~\ref{plot-models-Mbar}.  Solar metallicity models are plotted
with a diamond ($Z=0.02$, 0.02, and 0.019). Filled circles ($\bullet$)
are ellipticals, and open circles ($\circ$) are lenticular galaxies and
spiral bulges. The average error bars are plotted in the upper left
corners. \label{plot-models-IKbar}}
\end{figure}

Figure~\ref{plot-models-IKbar} compares the SSP models with
\Ibar$-$\Ksbar\ SBF color measurements, which are independent of the
galaxy distances.  Qualitatively the results are the same: the Liu
\etal\ and Worthey models give a large spread in age with roughly solar
metallicity (as in the case for \MbarKs, the errors in the SBF colors
allow for some spread in the metallicity); the Blakeslee \etal\ models
find a comparable spread in age with a larger spread in metallicity.
These comparisons using \Ibar--\Ksbar\ lead to slightly larger age
spreads and slightly smaller metallicity spreads than if we use \MbarKs.
The fluctuation colors for all the models follow the same trend,
becoming redder primarily as the metallicity increases.  The
Blakeslee~\etal\ models for \Ibar--\Ksbar\ are more similar to the other
models and better behaved than for \MbarKs; this further suggests there
may be a problem with their IR fluctuation magnitudes.

\citet{2000MNRAS.315..184K} has recently studied the star formation
history of early-type galaxies in the Fornax cluster using optical
absorption lines. He uses Balmer and metal lines to infer galaxian ages
and metallicities, respectively.  (See \citealp{tra00-2} for an
independent analysis of the same data set.)  There are 13~galaxies in
common between our work and his, most of which are ellipticals and not
S0's.  The faintest S0's in his sample, which show a very large spread
in age, are not in our SBF sample. The size of the age and metallicity
spreads inferred from his absorption line analysis and our SBF data are
in reasonable agreement.  He infers a range in [Fe/H] of --0.25 to
+0.25, compatible with our data especially given its errors. He finds an
age range of 5--12~Gyr, which is a slightly smaller spread and a higher
mean age than our results --- neither of these differences is surprising
given that the SBF data are more sensitive to recent star formation (see
below).

The above comparisons use single-burst population models.  But since the
star formation history of early-type galaxies remains uncertain, it is
important to consider the effect of multiple episodes of star formation.
In low redshift clusters, these galaxies follow a color-magnitude
relation with little intrinsic scatter, which is typically interpreted
as a small spread in age \citep{1992MNRAS.254..589B}. However,
intermediate-age (few Gyr) stellar populations can exist in small
amounts provided the bulk of the stars are old
\citep{1998MNRAS.299.1193B}.  Indeed, there are several lines of
evidence from distant ($z\gtrsim0.3$) clusters that early-type galaxies
have experienced more than a single episode of star formation
\citep[e.g.,][]{1994ApJ...432..453C, 1996MNRAS.279....1B,
1999ApJ...518..576P, ferreras01}, with the majority of the stellar mass
forming at high redshift but a minority fraction originating more
recently.

This evidence is based on optical colors and absorption line
measurements, which have several intrinsic differences from SBF data for
stellar populations.  Absorption lines are typically measured only in
the central regions of galaxies, while SBF measurements sample a much
larger volume.  Also, the light from main-sequence stars contribute to
the optical colors and absorption lines \citep[e.g.,
][]{1993ApJ...405..538B}, but the optical/IR SBF signal is much more
dependent on the giant stars. Because of this weighting to luminous cool
stars, changes in metallicity might be more easily discerned from SBF
measurements than from integrated spectral properties \citep{bc2000sbf}.
Also, the difference in weighting means that model predictions for
optical colors/lines and SBFs are subject to very different
uncertainties.  For instance, the interpretation of metal absorption
lines is hampered by the enhancement of $\alpha$-element in ellipticals
relative to the models \citep{1992ApJ...398...69W} and the limited range
in stellar temperatures and gravities used for model inputs
\citep[e.g.,][]{1994ApJS...94..687W, 2000ApJ...541..126M}, neither of
which have a direct bearing on SBF predictions.  Hence, concordance
between these lines of evidence would help confirm the robustness of the
results.

In a scenario where most of the stellar mass forms at high redshift with
a small fraction forming more recently, we would expect the
following. For recent star formation occurring with the same metallicity
as the older {\it in situ} population, galaxies will move in the \{\VmI,
\MbarKs\} plane along lines of constant metallicity.  Thus, the spread
in age inferred from the Liu \etal\ and Worthey models could arise from
late bursts of star formation with a common metallicity.  On the other
hand, if a late burst occurs with a different metallicity, we expect
noticeable changes to the position of the galaxies in the \{\VmI,
\MbarKs\} plane.  Despite its relatively small mass, the younger
population will be a substantial contributor to the total luminosity for
the first few~Gyr because of its much higher light-to-mass ratio.  Since
SBFs are sensitive to the second moment of the stellar luminosity
function, we expect the SBFs to be dominated by this younger population.
Ages inferred from SBFs would be expected to be younger and have larger
scatter than those from integrated colors/lines. Models predict that the
IR SBFs depend on metallicity (Figures~\ref{plot-models-Mbar}
and~\ref{plot-models-IKbar}), albeit to varying degrees.  The Worthey
and Liu \etal\ models predict a strong metallicity dependence at all
ages; if correct, then in the first few~Gyr after the burst, the IR SBFs
could be {\em either} brighter or fainter, depending on the metallicity
of the burst relative to the underlying population.

\begin{figure}
\centering
\hbox{\hskip -0.3in
\includegraphics[width=5in,angle=90]{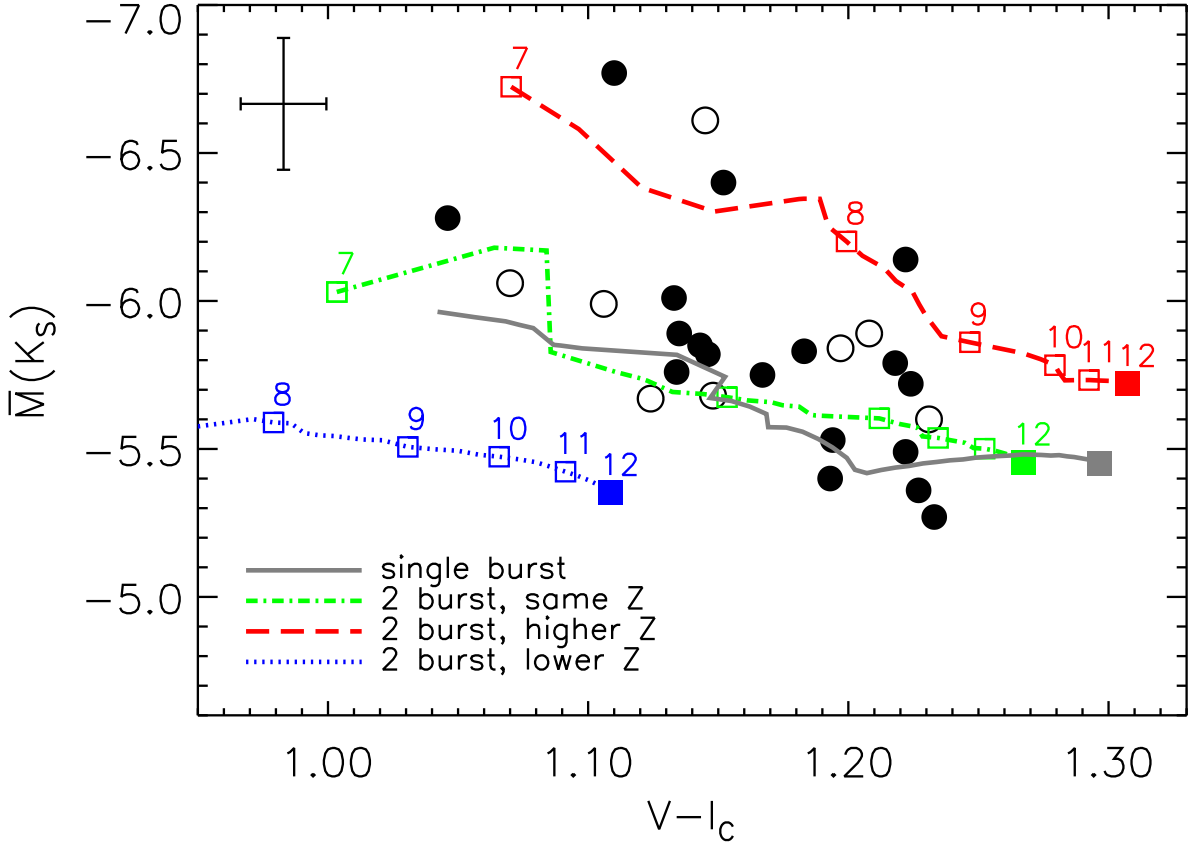}}
\caption{\normalsize An illustration of the effect of recent star
formation on the \Ks-band SBF magnitudes and integrated galaxy colors
using the stellar population models of \citet{bc2000sbf}.  Filled
circles ($\bullet$) are data for ellipticals, and open circles ($\circ$)
are for lenticular galaxies. The average error bars are plotted in the
upper left corner. The solid line shows a single-burst solar-metallicity
galaxy model aging from 2 to 12~Gyr, moving from left to right on the
plot.  The broken lines show models where a second burst of star
formation occurs 6~Gyr after the first burst and amounts to 20\% of the
final mass.  When the second burst has the same metallicity as the first
burst ({\em dot-dashed line}), the effect of the burst is to shift the
model-inferred age roughly along the same locus.  When the second burst
has a higher metallicity ([Fe/H]=+0.4; {\em upper dashed line}), the
\MbarKs\ is greatly increased while the range of \VmI\ galaxy color is
unaffected.  When the second burst has a lower metallicity
([Fe/H]=$-$1.7; {\em lower dotted line}), the \MbarKs\ is greatly
reduced.  The open squares are spaced 1~Gyr apart, starting at 7~Gyr age
(\ie, 1~Gyr after the burst occurs). The oldest (12~Gyr) version of each
model is marked with a filled square. \label{plot-burst2z}}
\end{figure}

Figure~\ref{plot-burst2z} illustrates the effect of recent star
formation using evolving population models of \citet{bc2000}.  The
majority of the model population has solar metallicity.  A second burst
of star formation is assumed to occur 6~Gyr after formation and
involving 20\% of the final galaxy mass. When the two bursts have the
same metallicity, the effect of the second burst is to reduce the
model-inferred age with little effect on the inferred metallicity.
However, when the second burst has a higher metallicity (in this case
[Fe/H]=+0.4), \MbarKs\ is greatly brightened while the range in \VmI\
galaxy color is unchanged.  Notice that the effect of the second burst
on \MbarKs\ lingers for many Gyr.  This is in sharp contrast to ordinary
optical colors, which revert to their pre-burst state in only a few~Gyr
\citep[e.g.,][]{1994ApJ...432..453C}. Similarly, for a second burst with
a much lower metallicity ([Fe/H]=$-$1.7, typical of local dwarf
galaxies), the range in optical colors is similar to the other models
only $\sim$2~Gyr after the burst, but the IR SBFs become much fainter
for long afterwards. Therefore, the offset in SBF magnitudes/colors
might be able to distinguish late bursts of star formation which have
very different metallicities from the main population, even after the
perturbations to the optical colors and absorption lines have subsided.

The two-burst models shown in Figure~\ref{plot-burst2z} should be taken
as illustrative, not definitive. They do show that if a small fraction
of the total galaxy mass was formed in the last few Gyr, this would
suffice to explain the spread of the optical/IR SBFs and integrated
colors.  In addition, the modeling suggests that the galaxies with
unusually bright \Ks-band SBFs, such as NGC~1389 and 1419, result from
recent star formation involving gas which was enriched in metals.
Likewise, the absence of galaxies with unusually {\em faint} \Ks-band
SBFs implies a lack of metal-poor intermediate-age stars, which would
form from unenriched infalling gas.  A more complete analysis of the
stellar populations of these galaxies is the subject of a future paper.


\section{Conclusions}
\label{sec:conclusions}

We have presented \Ks-band SBF data for 19 early-type Fornax cluster
galaxies.  This doubles the number of high-quality IR SBF measurements
to date.  Combining our measurements with data from the literature for
ellipticals in nearby clusters, we have calibrated \MbarKs\ as a
distance indicator.  We offer calibrations using either \HST\ Cepheid
distances to the clusters or $I$-band SBF distances to individual
galaxies.  When using the latter, we account for the covariance between
\VmI\ and \MbarKs. For both options, any systematic change in the \HST\
Cepheid distance scale will directly impact the \MbarKs\ calibration.

We have found \Ks-band fluctuation magnitudes vary considerably between
galaxies. However, we also find that the \Ks-band SBFs correlate with
\VmI\ galaxy color, like $I$-band SBFs.  The existence of this
correlation means \Ks-band distances can be improved by optical color
measurements to correct for stellar population variations between
galaxies.  It also suggests that the \Ho\ measurement by \citet{jen01}
using IR SBFs might be underestimated by $\sim$10\% since they assumed
no color-dependence for the SBF magnitudes.  The later-type galaxies may
follow a different correlation than the ellipticals, though this is
uncertain with the existing data.  In addition, this finding resolves
published concerns that the SBFs of NGC~4489 and M32 have anomalously
bright $K$-band SBFs: their brighter fluctuations are in accord with
their bluer integrated colors.  Overall, the intrinsic scatter in
$K$-band SBF appears to be significantly larger than $I$-band SBF.

We also find a few galaxies have very bright \Ks-band fluctuations, more
than 0.5~mag above the observed mean relation.  Most of these galaxies
are bluer and much less luminous than the bright early-type galaxies
used for cosmological distance measurements.  Nevertheless the existence
of such galaxies may be a concern for attempts to measure distances and
determine \Ho\ with IR SBFs.  Larger samples are needed to determine the
prevalence and extent of this phenomenon.

The existence of a correlation in \MbarKs\ (and \Ibar--\Ksbar) with
\VmI\ galaxy color is a new clue into the star formation histories of
cluster galaxies.  Such a relation is known to exist for $I$-band
SBFs. However, stellar population models predict the effects of age and
metallicity are largely degenerate in this bandpass, and hence $I$-band
SBFs alone are expected to be of little use for stellar population
studies.  On the other hand, models predict strong effects from stellar
population variations on IR fluctuation magnitudes and optical/IR
fluctuation colors.  Hence, our discovery of a {\em systematic} relation
between IR SBFs and galaxy color means we can determine specific
age/metallicity combinations, which then will be a reflection on the
star formation histories.  Given the current uncertainties in modeling
the RGB and AGB stars which dominate the IR SBF signal, these
interpretations will depend on the choice of stellar population models.
We argue the SBF model predictions from \citet{bc2000sbf} are probably
most reasonable based on what is known from resolved stellar populations
in Local Group star clusters. We also point out some potential
shortcomings with the models of \citet{bla00}.

Both the Liu \etal\ and \citet{1994ApJS...95..107W} models suggest that
the trend in \MbarKs\ with \VmI\ originates from variations in the age
of the populations dominating the SBF signal. Most of the galaxies are
inferred to have roughly solar metallicity, with perhaps a spread of a
factor of two.  Similar conclusions are reached using \Ibar--\Ksbar\
fluctuation colors, which are independent of galaxy distances.  All the
models suggest the single-burst equivalent ages for the bluest galaxies
are intermediate-age ($\lesssim$5~Gyr), implying these galaxies have
very luminous extended giant branches, similar to those found in
intermediate-age Magellanic Cloud clusters.  In the context of scenarios
where star formation occurs in more than one episode, the SBF/galaxy
color trend can be explained by the occurrence of late bursts of star
formation with metallicity similar to the older {\it in situ} population
comprising most of the stellar mass.  Such bursts act to change the
optical colors, which reduces the inferred age, without causing the
fluctuations to deviate from the main trend.

In a similar fashion, the few galaxies with very bright IR SBFs might be
explained by late bursts of star formation with a higher metallicity
than the old stars.  In this picture, the lack of galaxies with much
fainter IR fluctuations disfavors secondary star formation from more
metal-poor gas. The existence of galaxies with very bright fluctuations
among the bluer, less luminous galaxies indicates that the star
formation histories of these galaxies were more heterogeneous than those
of the redder, more massive galaxies.  This finding may seem at first to
be at variance with scenarios of hierarchical galaxy formation in a cold
dark matter dominated cosmology \citep[e.g.][]{1996MNRAS.283.1361B,
1998MNRAS.294..705K}, which predict that the more massive galaxies have
more extended formation histories.  However, in hierarchical cosmologies
merging rates depend strongly on mass --- low-mass galaxies experience
fewer recent mergers than high-mass galaxies. If mergers trigger
efficient star formation, \citet{kau01} find that lower-mass galaxies
will experience much more widely spaced bursts than massive galaxies,
and hence their star formation histories will be more heterogenous.

We caution that our results for the calibration of \Ks-band SBFs and the
inferred stellar populations are for galaxies in nearby clusters.  It
remains an open question how these results depend on environment.  Since
IR SBFs are predicted by all models to strongly depend on age and
metallicity, they could vary substantially with environment, though the
data for Local Group galaxies seem to agree well with those for galaxies
in the much denser Virgo and Fornax clusters.  Expanded samples will be
valuable both for strengthening the calibration of IR SBFs as
cosmological distance indicators and for deciphering the stellar
population histories of early-type galaxies.




\acknowledgments

It is a pleasure to acknowledge the CTIO staff for their support and
expert assistance in obtaining the observations presented here. In
particular, we thank Ron Probst, Bob Blum, Hernan Tirado, Patricio
Ugarte, Alberto Zu\~niga, and Patrice Bouchet.  M. Liu is also grateful
to NOAO/CTIO for providing travel support to carry out the observations
and for research support from the Beatrice Watson Parrent Fellowship at
the University of Hawai`i.  We have benefited from discussions with John
Tonry, Joe Jensen, Ed Ajhar, John Blakeslee, and Emory Bunn about
measuring SBFs. We also thank John Tonry and his collaborators for
providing their Fornax optical data in advance of publication.  We are
grateful to Andy Bunker for fruitful conversation on flat-fielding and
also to Eliot Malamuth for providing the seed of what became our IDL
ellipse fitting code.  This research has made use of the NASA/IPAC
Extragalactic Database (NED), the VizieR Service
\citep{2000A&AS..143...23O} at Centre de Donn\'ees astronomiques de
Strasbourg, the Digitized Sky Survey produced at the Space Telescope
Science Institute, and NASA's Astrophysics Data System Abstract
Service. This research was supported in part by grants to the authors
from the National Science Foundation (AST-9617173) and from NASA (\HST\
grant GO-07458.01-96A).


\appendix
\section{Revised SBF Models} \label{appendix}

In the version of the Bruzual-Charlot models used here, the thermally
pulsing phase of the asymptotic giant branch (TP-AGB) has been slightly
improved relative to the version used in \citet{bc2000sbf}.  The models
rely on the same calculations and spectral calibrations of TP-AGB stars
(M-type, C-type, and superwind phase) as described by Liu \etal\
However, the sampling of these stars has been improved to better account
for the extended giant branches of Magellanic Cloud and Galactic Bulge
clusters, as observed by \citet{1995MNRAS.272..391F} and
\citet{1998A&A...331...70G}.  Noticeable changes occur in all the SBF
magnitudes for the young (1--3~Gyr) models with sub-solar metallicities
and also in the optical SBFs (\Vbar~\Rbar~\Ibar) for the metal-rich
($Z\ge\Zsun$) models. Table~\ref{table-salp} presents the new set of
default models, which supersedes Table~2 of Liu \etal

\setcounter{section}{0}

\clearpage




\clearpage

%

\begin{deluxetable}{llcccccc}
\tablecaption{Fornax Cluster Galaxy Sample \label{table-gals}}
\tablewidth{0pt}

\tablehead{
\colhead{Name}   & 
\colhead{Type\tablenotemark{a}}   & 
\colhead{T\tablenotemark{b}}      & 
\multicolumn{1}{c}{$B_T$\tablenotemark{b}}  &
\multicolumn{1}{c}{Dia.\tablenotemark{a}} & 
\multicolumn{1}{c}{$\sigma$\tablenotemark{c}} &
\multicolumn{1}{c}{$A_B$\tablenotemark{d}}     & 
\colhead{$V-\Ic$\tablenotemark{e}} \\
\colhead{}   & 
\colhead{}   & 
\colhead{}      & 
\colhead{(mag)}  &
\colhead{(arcmin)} & 
\colhead{(km s\perone)} &
\colhead{(mag)}     & 
\colhead{(mag)} 
}

\startdata

IC 1919    &  SA(rs)0-- &  --3  &  13.80  &  1.6 $\times$ 1.2  &   26  &  0.06  &  1.108 $\pm$ 0.018  \\
IC 2006    &  E         &  --4  &  12.21  &  2.1 $\times$ 1.8  &  122  &  0.05  &  1.183 $\pm$ 0.018  \\
NGC 1336   &  SA0--     &  --3  &  13.10  &  2.1 $\times$ 1.5  &  115  &  0.05  &  1.124 $\pm$ 0.032  \\
NGC 1339   &  E4        &  --4  &  12.51  &  1.9 $\times$ 1.4  &  171  &  0.06  &  1.134 $\pm$ 0.012  \\
NGC 1344\tablenotemark{f} &  E5        &  --5  &  11.27  &  6.0 $\times$ 3.5  &  159  &  0.08  &  1.135 $\pm$ 0.011  \\
NGC 1351   &  SA0-- pec &  --3  &  12.46  &  2.8 $\times$ 1.7  &  147  &  0.06  &  1.148 $\pm$ 0.016  \\
NGC 1366   &  S0$^0$    &  --2  &  12.86  &  2.1 $\times$ 0.9  &\nodata&  0.07  &  1.095 $\pm$ 0.018  \\
NGC 1373   &  E         &  --4  &  14.12  &  1.1 $\times$ 0.9  &   79  &  0.06  &  1.085 $\pm$ 0.013  \\
NGC 1374   &  E         &  --4  &  12.00  &  2.5 $\times$ 2.3  &  207  &  0.06  &  1.146 $\pm$ 0.016  \\
NGC 1375   &  SAB0$^0$  &  --2  &  13.18  &  2.2 $\times$ 0.9  &   69  &  0.06  &  1.070 $\pm$ 0.019  \\
NGC 1379   &  E         &  --5  &  11.80  &  2.4 $\times$ 2.3  &  130  &  0.05  &  1.143 $\pm$ 0.019  \\
NGC 1380   &  SA0       &  --2  &  10.87  &  4.8 $\times$ 2.3  &  240  &  0.08  &  1.197 $\pm$ 0.019  \\
NGC 1380B\tablenotemark{g} &  SAB(s)0-- &  --3  &  13.87  &  1.5 $\times$ 1.3  &   99  &  0.07  &  1.106 $\pm$ 0.013  \\
NGC 1387   &  SAB(s)0-- &  --3  &  11.68  &  2.8 $\times$ 2.8  &\nodata&  0.06  &  1.208 $\pm$ 0.047  \\
NGC 1389   &  SAB(s)0   &  --3  &  12.42  &  2.3 $\times$ 1.4  &\nodata&  0.05  &  1.145 $\pm$ 0.019  \\
NGC 1399   &  E1 pec    &  --5  &  10.55  &  6.9 $\times$ 6.5  &  359  &  0.06  &  1.227 $\pm$ 0.016  \\
NGC 1404   &  E1        &  --5  &  10.97  &  3.3 $\times$ 3.0  &  242  &  0.05  &  1.224 $\pm$ 0.016  \\
NGC 1419   &  E         &  --5  &  13.48  &  1.1 $\times$ 1.1  &  129  &  0.06  &  1.110 $\pm$ 0.018  \\
NGC 1427   &  E5        &  --4  &  11.77  &  3.6 $\times$ 2.5  &  170  &  0.05  &  1.152 $\pm$ 0.018  \\

\enddata



\tablenotetext{a}{From the NASA/IPAC Extragalactic Database (NED).}

\tablenotetext{b}{From \citet{RC3} (RC3).}

\tablenotetext{c}{Data are from Hypercat ({\tt
http://www-obs.univ-lyon1.fr/hypercat}), which uses an updated version
of the literature compilation by \citet{1996A&A...309..749P}. The
exception is NGC~1373; its datum is from \citet{1998A&AS..133..325G}.}

\tablenotetext{d}{From \citet{1998ApJ...500..525S}.}

\tablenotetext{e}{From \citet{sbf4}.}

\tablenotetext{f}{a.k.a.\ NGC 1340.}

\tablenotetext{g}{a.k.a.\ NGC 1382.}


\end{deluxetable}


\begin{deluxetable}{lcccccc}
\tablecaption{\Ks-band SBF Measurements \label{table-sbfmags}}
\tablewidth{0pt}
\tabletypesize{\small}

\tablehead{
\colhead{Galaxy}          & 
\multicolumn{1}{c}{$\langle\mu_{K_S}\rangle$} & 
\colhead{$(P_0\!-\!P_r)/P_1$}       &
\multicolumn{1}{c}{$\mbar_{K_S}$} &
\colhead{$\Mbar_{K_S}^{I\hbox{-}SBF}$} &
\colhead{$\Mbar_{K_S}^{Cepheid}$}  &
\colhead{$\Ibar-\Ksbar$}  \\
\colhead{}  & 
\colhead{(mag~arcsec\pertwo)} & 
\colhead{}  & 
\colhead{(mag)} &
\colhead{(mag)} & 
\colhead{(mag)} &
\colhead{(mag)}  
}

\startdata

%
%

    IC 1919  &  18.52  &  \phn6.0 $\pm$ 2.5\phs &  24.84 $\pm$ 0.50  &  $-$6.47 $\pm$ 0.53  &  $-$6.58 $\pm$ 0.50  &  4.54 $\pm$ 0.52  \\
    IC 2006  &  17.77  &  \phn4.7 $\pm$ 0.3\phs &  25.76 $\pm$ 0.11  &  $-$5.83 $\pm$ 0.31  &  $-$5.66 $\pm$ 0.14  &  4.23 $\pm$ 0.29  \\
   NGC 1336  &  18.49  &  \phn2.9 $\pm$ 0.4\phs &  25.71 $\pm$ 0.23  &  $-$5.67 $\pm$ 0.31  &  $-$5.71 $\pm$ 0.25  &  3.82 $\pm$ 0.28  \\
   NGC 1339  &  17.57  &  \phn4.3 $\pm$ 0.4\phs &  25.85 $\pm$ 0.08  &  $-$5.76 $\pm$ 0.36  &  $-$5.57 $\pm$ 0.12  &  3.94 $\pm$ 0.36  \\
   NGC 1344  &  17.21  &     18.6 $\pm$ 0.9\phs &  25.59 $\pm$ 0.10  &  $-$5.89 $\pm$ 0.31  &  $-$5.83 $\pm$ 0.12  &  4.08 $\pm$ 0.31  \\
   NGC 1351  &  17.60  &  \phn5.4 $\pm$ 0.6\phs &  25.93 $\pm$ 0.11  &  $-$5.68 $\pm$ 0.19  &  $-$5.48 $\pm$ 0.13  &  3.93 $\pm$ 0.17  \\
   NGC 1366  &  19.00  &  \phn2.2 $\pm$ 0.6\phs &  25.51 $\pm$ 0.43  &  $-$6.11 $\pm$ 0.52  &  $-$5.91 $\pm$ 0.43  &  4.12 $\pm$ 0.51  \\
   NGC 1373  &  18.56  &  \phn9.0 $\pm$ 1.9\phs &  25.11 $\pm$ 0.37  &  $-$6.67 $\pm$ 0.60  &  $-$6.31 $\pm$ 0.38  &  4.63 $\pm$ 0.59  \\
   NGC 1374  &  17.05  &  \phn8.3 $\pm$ 0.3\phs &  25.66 $\pm$ 0.05  &  $-$5.82 $\pm$ 0.14  &  $-$5.75 $\pm$ 0.09  &  4.06 $\pm$ 0.11  \\
   NGC 1375  &  17.82  &     10.6 $\pm$ 1.2\phs &  25.52 $\pm$ 0.26  &  $-$6.06 $\pm$ 0.29  &  $-$5.90 $\pm$ 0.27  &  3.96 $\pm$ 0.28  \\
   NGC 1379  &  17.08  &     12.0 $\pm$ 0.9\phs &  25.66 $\pm$ 0.08  &  $-$5.85 $\pm$ 0.17  &  $-$5.75 $\pm$ 0.12  &  4.08 $\pm$ 0.14  \\
   NGC 1380  &  16.28  &  \phn7.4 $\pm$ 0.5\phs &  25.39 $\pm$ 0.04  &  $-$5.84 $\pm$ 0.18  &  $-$6.02 $\pm$ 0.09  &  4.31 $\pm$ 0.15  \\
   NGC 1380B &  17.95  &  \phn5.3 $\pm$ 0.4\phs &  25.78 $\pm$ 0.25  &  $-$5.99 $\pm$ 0.40  &  $-$5.63 $\pm$ 0.27  &  4.05 $\pm$ 0.39  \\
   NGC 1387  &  17.28  &  \phn8.2 $\pm$ 0.7\phs &  25.65 $\pm$ 0.06  &  $-$5.89 $\pm$ 0.27  &  $-$5.77 $\pm$ 0.10  &  4.41 $\pm$ 0.15  \\
   NGC 1389  &  17.53  &     14.0 $\pm$ 0.8\phs &  25.07 $\pm$ 0.10  &  $-$6.61 $\pm$ 0.21  &  $-$6.35 $\pm$ 0.13  &  4.85 $\pm$ 0.19  \\
   NGC 1399  &  16.01  &  \phn5.2 $\pm$ 0.5\phs &  26.14 $\pm$ 0.12  &  $-$5.36 $\pm$ 0.20  &  $-$5.28 $\pm$ 0.15  &  3.97 $\pm$ 0.18  \\
   NGC 1404  &  16.67  &  \phn8.9 $\pm$ 0.6\phs &  25.89 $\pm$ 0.05  &  $-$5.72 $\pm$ 0.20  &  $-$5.53 $\pm$ 0.10  &  4.31 $\pm$ 0.17  \\
   NGC 1419  &  17.92  &     16.6 $\pm$ 1.4\phs &  24.65 $\pm$ 0.13  &  $-$6.77 $\pm$ 0.27  &  $-$6.77 $\pm$ 0.15  &  4.85 $\pm$ 0.26  \\
   NGC 1427  &  17.51  &  \phn9.5 $\pm$ 0.5\phs &  25.46 $\pm$ 0.06  &  $-$6.40 $\pm$ 0.25  &  $-$5.96 $\pm$ 0.10  &  4.67 $\pm$ 0.23  \\

\enddata



\tablecomments{The quantity $(P_0\!-\!P_r)/P_1$ measures the S/N of the
stellar SBF signal (\ie, includes correction for unresolved
contaminating point sources).  The tabulated $\mbar_{K_S}$ has been
corrected for extinction. \Ks-band \Mbar's were computed using two
distance measurements to the galaxies: $I$-band SBF distances to
individual galaxies and a Cepheid distance to the cluster as a whole.
The errors on the $\Mbar_{K_S}$ from the Cepheid group distance includes
an allowance for the cluster's rms depth but {\em does not} include the
systematic error in the \HST\ Cepheid zero point, estimated to be
$\pm$0.16~mag \citep{2000ApJ...529..786M}.}

\end{deluxetable}

%
%

\begin{deluxetable}{llccccc}
\tablecaption{Calibration of \MbarKs \label{table-calibrate}}
\tablewidth{0pt}

\tablehead{
\colhead{Clusters}   & 
\colhead{Distances}   & 
\colhead{$a$}   & 
\colhead{$b$}   & 
\colhead{$N$}   &
\colhead{rms}   &
\colhead{$\tilde\chi^2$}
}

\startdata

Fornax only  & $I$-band SBF (individual) &  $-$5.81 $\pm$ 0.06  &  \nodata        &  14  &  0.23  &  1.0   \\
             &                           &  $-$5.84 $\pm$ 0.06  &  2.0 $\pm$ 1.4  &  14  &  0.21  &  0.9   \\ 
             & Cepheid (cluster)         &  $-$5.74 $\pm$ 0.03  &  \nodata        &  14  &  0.20  &  2.8   \\
             &                           &  $-$5.78 $\pm$ 0.04  &  4.1 $\pm$ 1.6  &  14  &  0.22  &  2.5   \\
&&&&&& \\
All galaxies & $I$-band SBF (individual)          &  $-$5.76 $\pm$ 0.04  &  \nodata        &  24  &  0.28  &  2.1  \\
             &                                    &  $-$5.84 $\pm$ 0.04  &  3.6 $\pm$ 0.8  &  24  &  0.22  &  1.2  \\ 
             & Cepheid (cluster)\tablenotemark{a} &  $-$5.68 $\pm$ 0.03  &  \nodata        &  23  &  0.25  &  4.2  \\
             &                                    &  $-$5.75 $\pm$ 0.03  &  4.3 $\pm$ 0.7  &  23  &  0.23  &  2.4  \\

\enddata



\tablenotetext{a}{Does not include NGC~1407 in Eridanus since no Cepheid
distance is available.}

\tablecomments{Calibration of \MbarKs\ using only our Fornax results and
our results combined with revised published data for Virgo, Leo,
Eridanus, and Local Group galaxies.  We either use $I$-band SBF
distances to individual galaxies or mean distances to the clusters from
Cepheid distances in \citet{2000ApJS..128..431F}.  The linear fits are
of the form \hbox{$\MbarKs = a + b\ [(V-\Ic)_0 - 1.15]$}, and the number
of galaxies used for the fit is tabulated ($N$).  Also listed are the
resulting rms (in mags) of the points after the fit and the reduced
chi-square ($\tilde\chi^2$) . See \S~\ref{sec:Mbarcalibrate} for
details.}

\end{deluxetable}


\thispagestyle{empty}
\begin{deluxetable}{ccccccccccccccccccccccccccc}
\rotate
\tablecaption{Revised BC2000 Model Predictions \label{table-salp}}
\tablewidth{0pt}
\tabletypesize{\scriptsize}
\setlength{\tabcolsep}{0.02in}

\tablenum{A1}

\tablehead{ \colhead{${Z}$} & \colhead{Gyr} & \colhead{$\overline{B}$} & \colhead{$\overline{V}$} & \colhead{$\overline{R_c}$} & \colhead{$\overline{I_c}$} & \colhead{$\overline{F814W}$} & \colhead{$\overline{F110M}$} & \colhead{$\overline{F110W}$} & \colhead{$\overline{J}$} & \colhead{$\overline{F160W}$} & \colhead{$\overline{H}$} & \colhead{$\overline{K^{\prime}}$} & \colhead{$\overline{K_s}$} & \colhead{$\overline{K}$} & \colhead{$\overline{F222M}$} & \colhead{$\overline{L}$} & \colhead{$\overline{L^{\prime}}$} & \colhead{$\overline{M}$} & \colhead{$V$--$I_c$} & \colhead{$V$--$K$} & \colhead{$J$--$K$} & \colhead{H$\beta$} & \colhead{Mg$_2$} & \colhead{Mgb} & \colhead{H$\gamma_A$} & \colhead{C4668}}


\startdata

0.0001 & $ 1$& $ 0.06$& $-0.56$& $-1.27$& $-2.24$& $-2.15$& $-3.24$& $-3.22$& $-3.66$& $-4.90$& $-5.07$& $-5.58$& $-5.60$& $-5.65$& $-5.57$& $-7.26$& $-7.29$& $-7.58$& $ 0.41$& $ 1.15$& $ 0.43$& $ 6.72$& $-0.04$& $-2.28$& \phs$ 8.79$& \phs$ 0.97$ \\
0.0001 & $ 2$& $ 0.35$& $-0.78$& $-1.66$& $-2.63$& $-2.54$& $-3.58$& $-3.56$& $-3.96$& $-5.11$& $-5.26$& $-5.77$& $-5.78$& $-5.82$& $-5.75$& $-6.83$& $-6.89$& $-7.10$& $ 0.69$& $ 1.77$& $ 0.60$& $ 4.72$& $-0.04$& $-1.31$& \phs$ 6.51$& \phs$ 0.38$ \\
0.0001 & $ 3$& $ 0.46$& $-0.68$& $-1.54$& $-2.47$& $-2.39$& $-3.39$& $-3.37$& $-3.77$& $-4.91$& $-5.06$& $-5.58$& $-5.59$& $-5.63$& $-5.56$& $-6.55$& $-6.61$& $-6.81$& $ 0.73$& $ 1.82$& $ 0.60$& $ 4.22$& $-0.03$& $-0.98$& \phs$ 6.07$& \phs$ 0.20$ \\
0.0001 & $ 5$& $ 0.46$& $-0.70$& $-1.50$& $-2.28$& $-2.22$& $-3.03$& $-3.00$& $-3.36$& $-4.37$& $-4.51$& $-5.06$& $-5.07$& $-5.12$& $-5.01$& $-6.03$& $-6.06$& $-6.32$& $ 0.78$& $ 1.85$& $ 0.57$& $ 3.48$& $-0.01$& $-0.47$& \phs$ 4.86$& $-0.08$ \\
0.0001 & $ 8$& $ 0.46$& $-0.66$& $-1.44$& $-2.14$& $-2.09$& $-2.69$& $-2.66$& $-2.93$& $-3.58$& $-3.66$& $-3.93$& $-3.94$& $-4.00$& $-3.95$& $-5.28$& $-5.29$& $-5.60$& $ 0.81$& $ 1.86$& $ 0.54$& $ 3.10$& \phs$ 0.00$& $-0.13$& \phs$ 3.92$& $-0.24$ \\
0.0001 & $12$& $ 0.50$& $-0.58$& $-1.33$& $-2.00$& $-1.95$& $-2.52$& $-2.49$& $-2.76$& $-3.38$& $-3.46$& $-3.70$& $-3.71$& $-3.75$& $-3.72$& $-4.89$& $-4.89$& $-5.18$& $ 0.85$& $ 1.93$& $ 0.55$& $ 2.91$& \phs$ 0.02$& \phs$ 0.09$& \phs$ 3.08$& $-0.33$ \\
0.0001 & $17$& $ 0.51$& $-0.56$& $-1.29$& $-1.94$& $-1.89$& $-2.45$& $-2.42$& $-2.68$& $-3.28$& $-3.36$& $-3.55$& $-3.57$& $-3.59$& $-3.58$& $-4.44$& $-4.45$& $-4.68$& $ 0.89$& $ 2.00$& $ 0.57$& $ 2.73$& \phs$ 0.03$& \phs$ 0.27$& \phs$ 2.38$& $-0.42$ \\[8pt]

0.0004 & $ 1$& $ 0.41$& $-0.63$& $-1.45$& $-2.33$& $-2.26$& $-3.21$& $-3.18$& $-3.57$& $-4.65$& $-4.80$& $-5.28$& $-5.29$& $-5.34$& $-5.27$& $-7.00$& $-7.04$& $-7.34$& $ 0.58$& $ 1.47$& $ 0.49$& $ 5.50$& $-0.02$& $-1.24$& \phs$ 6.51$& $-0.39$ \\
0.0004 & $ 2$& $ 0.27$& $-1.03$& $-1.92$& $-2.83$& $-2.75$& $-3.71$& $-3.69$& $-4.07$& $-5.14$& $-5.29$& $-5.77$& $-5.78$& $-5.81$& $-5.75$& $-6.79$& $-6.84$& $-7.06$& $ 0.77$& $ 1.91$& $ 0.62$& $ 4.08$& $-0.01$& $-0.52$& \phs$ 5.09$& $-0.47$ \\
0.0004 & $ 3$& $ 0.47$& $-0.82$& $-1.68$& $-2.58$& $-2.50$& $-3.45$& $-3.43$& $-3.82$& $-4.90$& $-5.04$& $-5.54$& $-5.55$& $-5.59$& $-5.52$& $-6.51$& $-6.56$& $-6.78$& $ 0.79$& $ 1.93$& $ 0.62$& $ 3.75$& $-0.00$& $-0.30$& \phs$ 4.76$& $-0.54$ \\
0.0004 & $ 5$& $ 0.51$& $-0.83$& $-1.70$& $-2.51$& $-2.45$& $-3.27$& $-3.25$& $-3.60$& $-4.57$& $-4.71$& $-5.19$& $-5.20$& $-5.24$& $-5.17$& $-6.06$& $-6.10$& $-6.33$& $ 0.84$& $ 1.98$& $ 0.61$& $ 3.13$& \phs$ 0.01$& \phs$ 0.15$& \phs$ 3.27$& $-0.58$ \\
0.0004 & $ 8$& $ 0.53$& $-0.79$& $-1.60$& $-2.30$& $-2.25$& $-2.88$& $-2.85$& $-3.13$& $-3.80$& $-3.89$& $-4.14$& $-4.15$& $-4.19$& $-4.16$& $-5.34$& $-5.35$& $-5.64$& $ 0.88$& $ 2.00$& $ 0.58$& $ 2.66$& \phs$ 0.03$& \phs$ 0.50$& \phs$ 1.81$& $-0.61$ \\
0.0004 & $12$& $ 0.55$& $-0.74$& $-1.54$& $-2.23$& $-2.19$& $-2.81$& $-2.77$& $-3.06$& $-3.71$& $-3.79$& $-4.01$& $-4.01$& $-4.03$& $-4.03$& $-4.90$& $-4.91$& $-5.12$& $ 0.91$& $ 2.05$& $ 0.59$& $ 2.56$& \phs$ 0.04$& \phs$ 0.64$& \phs$ 1.53$& $-0.64$ \\
0.0004 & $17$& $ 0.55$& $-0.72$& $-1.52$& $-2.21$& $-2.16$& $-2.78$& $-2.75$& $-3.03$& $-3.68$& $-3.77$& $-3.96$& $-3.96$& $-3.97$& $-3.98$& $-4.65$& $-4.66$& $-4.80$& $ 0.93$& $ 2.10$& $ 0.60$& $ 2.64$& \phs$ 0.05$& \phs$ 0.66$& \phs$ 1.77$& $-0.66$ \\[8pt]

0.0040 & $ 1$& $ 1.20$& $-0.12$& $-1.33$& $-2.71$& $-2.60$& $-3.99$& $-3.97$& $-4.44$& $-5.76$& $-5.93$& $-6.55$& $-6.55$& $-6.61$& $-6.47$& $-7.24$& $-7.26$& $-7.55$& $ 0.73$& $ 2.01$& $ 0.74$& $ 4.75$& \phs$ 0.05$& \phs$ 0.10$& \phs$ 6.52$& \phs$ 0.04$ \\
0.0040 & $ 2$& $ 1.32$& $-0.27$& $-1.36$& $-2.49$& $-2.40$& $-3.55$& $-3.51$& $-3.92$& $-5.09$& $-5.25$& $-5.88$& $-5.89$& $-5.96$& $-5.82$& $-6.85$& $-6.86$& $-7.23$& $ 0.91$& $ 2.28$& $ 0.76$& $ 3.30$& \phs$ 0.08$& \phs$ 1.03$& \phs$ 3.12$& \phs$ 0.72$ \\
0.0040 & $ 3$& $ 1.40$& $-0.19$& $-1.26$& $-2.38$& $-2.29$& $-3.41$& $-3.37$& $-3.75$& $-4.87$& $-5.02$& $-5.61$& $-5.62$& $-5.68$& $-5.55$& $-6.47$& $-6.47$& $-6.83$& $ 1.00$& $ 2.42$& $ 0.77$& $ 2.63$& \phs$ 0.10$& \phs$ 1.57$& \phs$ 0.42$& \phs$ 1.27$ \\
0.0040 & $ 5$& $ 1.64$& \phs$ 0.06$& $-1.04$& $-2.23$& $-2.13$& $-3.27$& $-3.21$& $-3.58$& $-4.66$& $-4.80$& $-5.35$& $-5.36$& $-5.40$& $-5.31$& $-6.09$& $-6.09$& $-6.38$& $ 1.04$& $ 2.47$& $ 0.77$& $ 2.39$& \phs$ 0.12$& \phs$ 1.85$& $-0.93$& \phs$ 1.37$ \\
0.0040 & $ 8$& $ 1.67$& \phs$ 0.16$& $-0.92$& $-2.17$& $-2.06$& $-3.27$& $-3.19$& $-3.54$& $-4.53$& $-4.67$& $-5.10$& $-5.11$& $-5.10$& $-5.13$& $-5.83$& $-5.84$& $-6.02$& $ 1.11$& $ 2.57$& $ 0.77$& $ 2.06$& \phs$ 0.14$& \phs$ 2.17$& $-2.65$& \phs$ 1.64$ \\
0.0040 & $12$& $ 1.76$& \phs$ 0.29$& $-0.79$& $-2.06$& $-1.96$& $-3.13$& $-3.05$& $-3.37$& $-4.31$& $-4.44$& $-4.81$& $-4.81$& $-4.78$& $-4.84$& $-5.19$& $-5.20$& $-5.05$& $ 1.15$& $ 2.63$& $ 0.77$& $ 1.84$& \phs$ 0.15$& \phs$ 2.43$& $-3.56$& \phs$ 1.80$ \\
0.0040 & $17$& $ 1.72$& \phs$ 0.35$& $-0.71$& $-1.96$& $-1.85$& $-3.00$& $-2.92$& $-3.24$& $-4.17$& $-4.30$& $-4.66$& $-4.66$& $-4.63$& $-4.69$& $-4.94$& $-4.96$& $-4.68$& $ 1.18$& $ 2.67$& $ 0.77$& $ 1.76$& \phs$ 0.17$& \phs$ 2.56$& $-3.77$& \phs$ 1.80$ \\[8pt]

0.0080 & $ 1$& $ 1.49$& \phs$ 0.29$& $-0.88$& $-2.39$& $-2.27$& $-3.83$& $-3.79$& $-4.28$& $-5.67$& $-5.85$& $-6.56$& $-6.57$& $-6.64$& $-6.47$& $-7.38$& $-7.39$& $-7.73$& $ 0.77$& $ 2.15$& $ 0.81$& $ 4.34$& \phs$ 0.07$& \phs$ 0.65$& \phs$ 5.78$& \phs$ 0.81$ \\
0.0080 & $ 2$& $ 1.76$& \phs$ 0.16$& $-0.96$& $-2.27$& $-2.16$& $-3.58$& $-3.49$& $-3.91$& $-5.08$& $-5.23$& $-5.89$& $-5.90$& $-5.97$& $-5.84$& $-6.84$& $-6.85$& $-7.23$& $ 0.99$& $ 2.51$& $ 0.82$& $ 2.95$& \phs$ 0.12$& \phs$ 1.76$& \phs$ 1.33$& \phs$ 2.05$ \\
0.0080 & $ 3$& $ 1.88$& \phs$ 0.27$& $-0.84$& $-2.15$& $-2.05$& $-3.46$& $-3.37$& $-3.78$& $-4.89$& $-5.04$& $-5.64$& $-5.65$& $-5.70$& $-5.61$& $-6.54$& $-6.55$& $-6.91$& $ 1.07$& $ 2.64$& $ 0.83$& $ 2.44$& \phs$ 0.15$& \phs$ 2.24$& $-1.24$& \phs$ 2.64$ \\
0.0080 & $ 5$& $ 2.20$& \phs$ 0.69$& $-0.41$& $-1.81$& $-1.69$& $-3.22$& $-3.11$& $-3.53$& $-4.63$& $-4.78$& $-5.35$& $-5.36$& $-5.39$& $-5.34$& $-6.17$& $-6.18$& $-6.51$& $ 1.09$& $ 2.65$& $ 0.82$& $ 2.27$& \phs$ 0.16$& \phs$ 2.49$& $-2.51$& \phs$ 2.68$ \\
0.0080 & $ 8$& $ 2.15$& \phs$ 0.75$& $-0.31$& $-1.76$& $-1.65$& $-3.38$& $-3.24$& $-3.69$& $-4.74$& $-4.90$& $-5.40$& $-5.42$& $-5.40$& $-5.46$& $-6.03$& $-6.06$& $-6.31$& $ 1.17$& $ 2.82$& $ 0.85$& $ 1.94$& \phs$ 0.18$& \phs$ 2.85$& $-4.06$& \phs$ 3.07$ \\
0.0080 & $12$& $ 2.18$& \phs$ 0.87$& $-0.16$& $-1.63$& $-1.52$& $-3.30$& $-3.16$& $-3.61$& $-4.64$& $-4.80$& $-5.26$& $-5.28$& $-5.24$& $-5.33$& $-5.75$& $-5.78$& $-5.96$& $ 1.23$& $ 2.92$& $ 0.86$& $ 1.70$& \phs$ 0.20$& \phs$ 3.13$& $-5.19$& \phs$ 3.32$ \\
0.0080 & $17$& $ 2.23$& \phs$ 0.98$& $-0.01$& $-1.48$& $-1.37$& $-3.13$& $-3.00$& $-3.46$& $-4.49$& $-4.65$& $-5.10$& $-5.11$& $-5.07$& $-5.16$& $-5.44$& $-5.47$& $-5.62$& $ 1.27$& $ 2.98$& $ 0.86$& $ 1.53$& \phs$ 0.21$& \phs$ 3.32$& $-5.79$& \phs$ 3.41$ \\[8pt]

0.0200 & $ 1$& $ 1.82$& \phs$ 0.69$& $-0.36$& $-1.89$& $-1.77$& $-3.61$& $-3.53$& $-4.05$& $-5.44$& $-5.62$& $-6.41$& $-6.42$& $-6.50$& $-6.32$& $-7.36$& $-7.36$& $-7.74$& $ 0.83$& $ 2.34$& $ 0.85$& $ 3.83$& \phs$ 0.12$& \phs$ 1.53$& \phs$ 4.25$& \phs$ 2.80$ \\
0.0200 & $ 2$& $ 2.31$& \phs$ 0.80$& $-0.26$& $-1.65$& $-1.56$& $-3.61$& $-3.43$& $-3.97$& $-5.07$& $-5.24$& $-5.90$& $-5.92$& $-5.96$& $-5.92$& $-6.89$& $-6.91$& $-7.31$& $ 1.04$& $ 2.74$& $ 0.87$& $ 2.76$& \phs$ 0.17$& \phs$ 2.47$& $-0.50$& \phs$ 4.01$ \\
0.0200 & $ 3$& $ 2.54$& \phs$ 1.00$& $-0.04$& $-1.50$& $-1.41$& $-3.72$& $-3.50$& $-4.08$& $-5.08$& $-5.26$& $-5.79$& $-5.83$& $-5.82$& $-5.91$& $-6.61$& $-6.69$& $-7.02$& $ 1.13$& $ 2.94$& $ 0.89$& $ 2.29$& \phs$ 0.20$& \phs$ 3.02$& $-3.33$& \phs$ 4.78$ \\
0.0200 & $ 5$& $ 2.76$& \phs$ 1.26$& \phs$ 0.24$& $-1.20$& $-1.11$& $-3.49$& $-3.26$& $-3.84$& $-4.84$& $-5.02$& $-5.55$& $-5.59$& $-5.57$& $-5.68$& $-6.24$& $-6.33$& $-6.65$& $ 1.18$& $ 3.00$& $ 0.89$& $ 2.01$& \phs$ 0.22$& \phs$ 3.35$& $-4.67$& \phs$ 5.08$ \\
0.0200 & $ 8$& $ 2.57$& \phs$ 1.35$& \phs$ 0.39$& $-1.01$& $-0.94$& $-3.38$& $-3.14$& $-3.74$& $-4.74$& $-4.92$& $-5.44$& $-5.48$& $-5.46$& $-5.58$& $-6.05$& $-6.16$& $-6.46$& $ 1.24$& $ 3.13$& $ 0.91$& $ 1.74$& \phs$ 0.25$& \phs$ 3.69$& $-5.82$& \phs$ 5.49$ \\
0.0200 & $12$& $ 2.47$& \phs$ 1.48$& \phs$ 0.57$& $-0.81$& $-0.75$& $-3.35$& $-3.09$& $-3.71$& $-4.70$& $-4.90$& $-5.42$& $-5.47$& $-5.45$& $-5.60$& $-5.95$& $-6.11$& $-6.40$& $ 1.30$& $ 3.26$& $ 0.93$& $ 1.51$& \phs$ 0.27$& \phs$ 4.00$& $-6.80$& \phs$ 5.85$ \\
0.0200 & $17$& $ 2.51$& \phs$ 1.57$& \phs$ 0.69$& $-0.67$& $-0.61$& $-3.30$& $-3.02$& $-3.67$& $-4.65$& $-4.86$& $-5.38$& $-5.44$& $-5.42$& $-5.58$& $-5.90$& $-6.10$& $-6.39$& $ 1.34$& $ 3.35$& $ 0.94$& $ 1.33$& \phs$ 0.29$& \phs$ 4.20$& $-7.50$& \phs$ 6.12$ \\[8pt]

0.0500 & $ 1$& $ 2.22$& \phs$ 1.08$& \phs$ 0.17$& $-1.16$& $-1.09$& $-4.45$& $-4.16$& $-5.03$& $-6.21$& $-6.43$& $-7.18$& $-7.24$& $-7.27$& $-7.33$& $-7.81$& $-7.96$& $-8.24$& $ 0.93$& $ 2.94$& $ 1.07$& $ 3.25$& \phs$ 0.18$& \phs$ 2.45$& \phs$ 1.62$& \phs$ 5.76$ \\
0.0500 & $ 2$& $ 2.80$& \phs$ 1.32$& \phs$ 0.32$& $-1.04$& $-0.97$& $-4.14$& $-3.86$& $-4.72$& $-5.77$& $-6.01$& $-6.65$& $-6.75$& $-6.77$& $-6.95$& $-7.35$& $-7.60$& $-7.92$& $ 1.14$& $ 3.23$& $ 1.01$& $ 2.42$& \phs$ 0.23$& \phs$ 3.32$& $-3.70$& \phs$ 7.08$ \\
0.0500 & $ 3$& $ 2.98$& \phs$ 1.48$& \phs$ 0.48$& $-0.82$& $-0.77$& $-3.85$& $-3.57$& $-4.39$& $-5.39$& $-5.62$& $-6.21$& $-6.31$& $-6.31$& $-6.51$& $-6.90$& $-7.15$& $-7.47$& $ 1.22$& $ 3.36$& $ 0.99$& $ 2.04$& \phs$ 0.27$& \phs$ 3.81$& $-5.52$& \phs$ 7.89$ \\
0.0500 & $ 5$& $ 2.78$& \phs$ 1.71$& \phs$ 0.75$& $-0.57$& $-0.52$& $-3.66$& $-3.38$& $-4.22$& $-5.21$& $-5.47$& $-6.02$& $-6.14$& $-6.14$& $-6.40$& $-6.62$& $-6.96$& $-7.29$& $ 1.31$& $ 3.52$& $ 1.02$& $ 1.71$& \phs$ 0.30$& \phs$ 4.32$& $-6.97$& \phs$ 8.58$ \\
0.0500 & $ 8$& $ 2.63$& \phs$ 1.87$& \phs$ 0.94$& $-0.36$& $-0.31$& $-3.52$& $-3.25$& $-4.11$& $-5.11$& $-5.38$& $-5.93$& $-6.06$& $-6.06$& $-6.35$& $-6.49$& $-6.88$& $-7.21$& $ 1.35$& $ 3.58$& $ 1.02$& $ 1.52$& \phs$ 0.32$& \phs$ 4.61$& $-7.68$& \phs$ 8.95$ \\
0.0500 & $12$& $ 2.42$& \phs$ 1.85$& \phs$ 0.98$& $-0.26$& $-0.22$& $-3.35$& $-3.08$& $-3.92$& $-4.92$& $-5.18$& $-5.73$& $-5.86$& $-5.86$& $-6.13$& $-6.28$& $-6.65$& $-6.99$& $ 1.41$& $ 3.69$& $ 1.02$& $ 1.32$& \phs$ 0.36$& \phs$ 4.96$& $-8.52$& \phs$ 9.69$ \\
0.0500 & $17$& $ 2.38$& \phs$ 1.91$& \phs$ 1.08$& $-0.10$& $-0.06$& $-3.14$& $-2.87$& $-3.71$& $-4.71$& $-4.97$& $-5.52$& $-5.64$& $-5.64$& $-5.90$& $-6.05$& $-6.41$& $-6.76$& $ 1.45$& $ 3.73$& $ 1.01$& $ 1.16$& \phs$ 0.38$& \phs$ 5.22$& $-9.01$& \phs$10.28$ \\
\enddata
\end{deluxetable}

\end{document}